\documentclass[twocolumn,trackchanges]{aastex62}

\received{XX}
\revised{YY}
\accepted{ZZ}

\submitjournal{ApJ}
\shorttitle{}
\shortauthors{Rosales, J. A.}

\usepackage{txfonts}
\newcommand{\kms}{\,km\,s$^{-1}$}

\begin{document}

\title{Stellar and accretion disk parameters of the close binary HD\,50526 }

\correspondingauthor{Rosales, J. A.}
\email{jrosales@astro-udec.cl}

\author[0000-0001-6657-9121]{Rosales, J. A.}
\affiliation{Department of Physics, North-West University, Private Bag X2046, Mmabatho 2735, South Africa.}
\affiliation{Departamento de Astronom\'{i}a, Universidad de Concepci\'{o}n, Casilla 160-C, Concepci\'{o}n, Chile.}
\affiliation{Main Astronomical Observatory, National Academy of Sciences of Ukraine, 27 Akademika Zabolotnoho St, 03680 Kyiv, Ukraine.}

\author{Mennickent, R. E.}
\affiliation{Departamento de Astronom\'{i}a, Universidad de Concepci\'{o}n, Casilla 160-C, Concepci\'{o}n, Chile.}

\author{Djura\v{s}evi\'{c}, G.}
\affiliation{Astronomical Observatory, Volgina 7, 11060 Belgrade 38, Serbia.}
\affiliation{Isaac Newton Institute of Chile, Yugoslavia Branch, 11060 Belgrade, Serbia.}

\author{Schleicher, D. R. G.}
\affiliation{Departamento de Astronom\'{i}a, Universidad de Concepci\'{o}n, Casilla 160-C, Concepci\'{o}n, Chile.}

\author{Zharikov, S.}
\affiliation{Instituto de Astronom\'{i}a, Universidad Nacional Aut\'{o}noma de M\'{e}xico, Apartado Postal 877, Ensenada, Baja California, 22800 M\'{e}xico.}

\author{Araya, I.}
\affiliation{Centro de Investigación DAiTA Lab, Facultad de Estudios Interdisciplinarios, Universidad Mayor, Chile}

\author{Celed\'{o}n, L.}
\affiliation{Instituto de F\'{i}sica y Astronom\'{i}a, Facultad de Ciencias, Universidad de Valpara\'{i}so, Chile.}

\author{Cur\'{e}, M.}
\affiliation{Instituto de F\'{i}sica y Astronom\'{i}a, Facultad de Ciencias, Universidad de Valpara\'{i}so, Chile.}

\begin{abstract}

We present a photometric and spectroscopic study of HD\,50526, an ellipsoidal binary member of the group Double Periodic Variable stars. Performing data-mining in photometric surveys and conducting new spectroscopic observations with several spectrographs during 2008 to 2015, we obtained orbital and stellar parameters of the system. The radial velocities were analyzed with the genetic \texttt{PIKAIA} algorithm, whereas Doppler tomography maps for the H$\alpha$ and H$\beta$ lines were constructed with the Total Variation Minimization code. An optimized simplex-algorithm was used to solve the inverse-problem adjusting the light curve with the best stellar parameters for the system. We find an  orbital period of $6\fd701 \pm 0\fd001$ and a long photometric cycle of $191 \pm 2 ~\mathrm{d}$. We detected the spectral features of the coldest star, and modeled it with a $\log{g} = 2.79 \pm 0.02 ~\mathrm{dex}$ giant of mass $1.13 \pm 0.02 ~\mathrm{M_{\odot}}$ and effective temperature $10500 \pm 125 ~\mathrm{K}$. In addition, we determine a mass ratio $q= 0.206 \pm 0.033$ and that the hot star is a B-type dwarf of mass $5.48 \pm 0.02 ~\mathrm{M_{\odot}}$. The $V$-band orbital  light curve can be modeled including the presence of an accretion disk around the hotter star. This fills the Roche lobe of the hotter star, and has a radius $14.74 \pm 0.02 ~\mathrm{R_{\odot}}$ and temperature at the outer edge 9400\,K. Two bright spots located in the disk account for the global morphology of the light curve. The Doppler tomography maps of H$\alpha$ and H$\beta$, reveal complex structures of mass fluxes in the system.
\end{abstract}


\section{Introduction}
\label{Sec: Sec. 1}

Since the discovery of the first Double Periodic Variable stars by \citet{2003A&A...399L..47M}, in the Large Magellanic Cloud (LMC) and the Small Magellanic Cloud (SMC), we have learn that there is a new kind of semi-detached, mass-transferring binary stars showing two closely linked photometric variations. These systems show an enigmatic long period on average 33 times longer than the orbital period \citep{2016MNRAS.455.1728M,2017SerAJ.194....1M,2010AcA....60..179P,2013AcA....63..323P}. But an interesting and remarkable property of the DPVs is the constancy of their orbital periods, which usually does not occur in the algols undergoing Roche Lobe Overflow  mass transfer \citep{2013MNRAS.428.1594G}. To date, it is suspected that some interacting binary systems show variations of the wind generated in the stream-disk impact region \citep{2016MNRAS.461.1674M, 2008A&A...487.1129V}, e.g., the interacting binary V393 Scorpii studied by \citet{2012MNRAS.427..607M} shows evidence of a cyclically variable bipolar wind. The prototype $\beta$ Lyrae also is a DPV, and shows evidence of a jet emanating from the the accretion disk in a process far to be well understood  \citep{1996A&A...312..879H}. 

Recently, the nature of the second photometric period in these stars was associated to a mechanism based on cycles of a magnetic dynamo in the donor star \citep{2017A&A...602A.109S,2018PASP..130i4203M}. Also, it is believed that the changes in the DPVs orbital light curves could be related to changes in disk size/temperature and spot temperature/position \citep{2018MNRAS.477L..11G}. Understanding the mechanism associated to the DPV phenomenon will be of fundamental importance for the study of the mass transfer in semidetached binaries related to algols. Also it can provide information about the stellar magnetic dynamos, stellar densities, tidal friction strength and wind processes involved in these systems, processes that have been considered in stellar population synthesis by \citet{2014ApJ...782....7D} and also studied by \citet{2015A&A...577A..55D}. 

The interacting binary HD\,50526 (ASAS ID $065402 + 0648.8$, $\alpha_{2000}= 06:54:02.0$, $\delta_{2000}= 06:48:47.9$, $V= 8.23 \pm 0.01 ~\mathrm{mag}$, $B - V= 0.08 ~\mathrm{mag}$, spectral type B9)\footnote{\url{http://simbad.u-strasbg.fr/simbad}}, is a system characterized by a long photometric cycle of 189\fd5  in the ASAS \footnote{\url{http://www.astrouw.edu.pl/asas/?page=acvs}} catalogue \citep{1997AcA....47..467P} and shows a short period of 6\fd7007, as published in VSX\footnote{\url{https://www.aavso.org/vsx/}}. The distance based on the GAIA\footnote{\url{http://gea.esac.esa.int/archive/}} DR2 parallax is 1247 [+50 -47] pc \citep{2018AJ....156...58B}. Since very few DPVs have been studied spectroscopically, this object requires a detailed study, specially aimed to determine its orbital and stellar parameters, and elucidate its enigmatic long period. In this paper we shed light on the physical parameters of this interesting system. 

In Section \ref{Sec: Sec. 2}  we present a detailed photometric analysis of ASAS photometry for HD\,50526 finding ephemeris for the orbital and long cycles. In Section \ref{Sec: Sec. 3} we present a summary of our spectroscopic data and our methods of data reduction. In Section \ref{Sec: Sec. 4} we use the method of spectral disentangling to separate the  stellar components and analyze them individually. We also obtain the orbital parameters of the system using a code based in a genetic algorithm and constrain the physical parameters of the donor through a minimization of the deviation between theoretical normalized spectra and the observed one. In this section we also calculate residual emission spectra and construct Doppler tomography maps for the H$\alpha$ and H$\beta$ lines. In Section \ref{Sec: Sec. 5} we model the light curve finding additional constrains for the stellar and orbital parameters and in Section \ref{Sec: Sec. 6} we discuss our results. Finally, the main results of our research are summarized in Section \ref{Sec: Sec. 7}.\\

\section{Photometric Ephemeris}
\label{Sec: Sec. 2}

The ASAS Photometric Catalog is maintained separately for each observed field, so for some stars independent data sets of measurements are available, their mean magnitudes may slightly differ. In addition, these data are labeled as A for the best data, B for mean data, C corresponds to A and B data and finally D are the worst data. The catalog calculates the magnitudes using different apertures, wherein the photometric data are labeled from Mag$_{0}$ to Mag$_{4}$ with steps of 1, these have size of 2 pixels for the smallest to 6 pixel for the largest aperture.  The smallest aperture is appropriate to the weakest stars, while the largest apertures are to the brightest stars. So to get rid of the saturated measurements, as an example it is optimal to compare Mag$_{0}$ and Mag$_{4}$ and if the difference between these is greater than 0.05, it is considered as saturation. Therefore, with this information we performed a photometric analysis of the ASAS light curve considering the 269 better-quality data points labeled as A-type by ASAS. 
The data set was analyzed with the Phase Dispersion Minimization algorithm \texttt{PDM-IRAF} \citep{1978ApJ...224..953S} \footnote{IRAF is distributed by National Optical Astronomy Observatories, which are operated by Association of Universities for Research in Astronomy, Inc., under cooperative agreement with the National Science Foundation}, revealing an orbital period $P_\mathrm{o}=6\fd701 \pm 0\fd001$. We also determined the epoch of minimum $\mathrm{HJD} = 2453096.5463$. The data set also revealed a long period $P_\mathrm{l}=191 \pm 2 ~\mathrm{d}$ with a maximum at $\mathrm{HJD}= 2452942.84607$, presenting a mean magnitude of the system of $V= 8.296 \pm 0.044 ~\mathrm{mag}$ and variations from $V_\mathrm{max}= 8.148 \pm 0.049 ~\mathrm{mag}$ to $V_\mathrm{min} = 8.615 \pm 0.045 ~\mathrm{mag}$ at the visual.

\begin{table*}
	\caption{Summary of ephemeris for HD\,50526. We give the orbital period ($P_\mathrm{o}$) and epoch for the minimum brightness ($\mathrm{T_{0}}$). 
		The initial and final HJDs of the photometric time series (minus -2450000) are given by $\mathrm{T_{i}}$ and $\mathrm{T_{f}}$. N indicates the number of data points. We give the mean $V$ value along with the standard deviation in parenthesis.}
	\label{Tab: Tab. 1}
	\centering
	\begin{tabular}{llcccccccc}
		\hline
		\hline
		\noalign{\smallskip}
		\textrm{Source} & \textrm{P$_\mathrm{o}$ (d)} & \textrm{P$_\mathrm{l}$ (d)} & \textrm{T$_{0}$\,(min$_{\mathrm{o}}$) } & \textrm{T$_{0}$\,(max$_{\mathrm{l}}$) }& \textrm{$V$ (mag)} & \textrm{N}& \textrm{T$_\mathrm{i}$}& \textrm{T$_\mathrm{f}$} \\ 
		\hline
		\noalign{\smallskip}
		\textrm{ASAS}		& \textrm{6.701(1) } & \textrm{191(2)} & \textrm{3096.54630}& \textrm{2942.84607}& \textrm{8.296(44)}& \textrm{269}& \textrm{2549.84189}& \textrm{5162.78517}\\
		\hline       
	\end{tabular}
	\vspace{0.25cm}
	\\
\end{table*}

\begin{figure}
	\begin{center}
		\includegraphics[trim=0.2cm 0.2cm 0.3cm 0.2cm,clip,width=0.45\textwidth,angle=0]{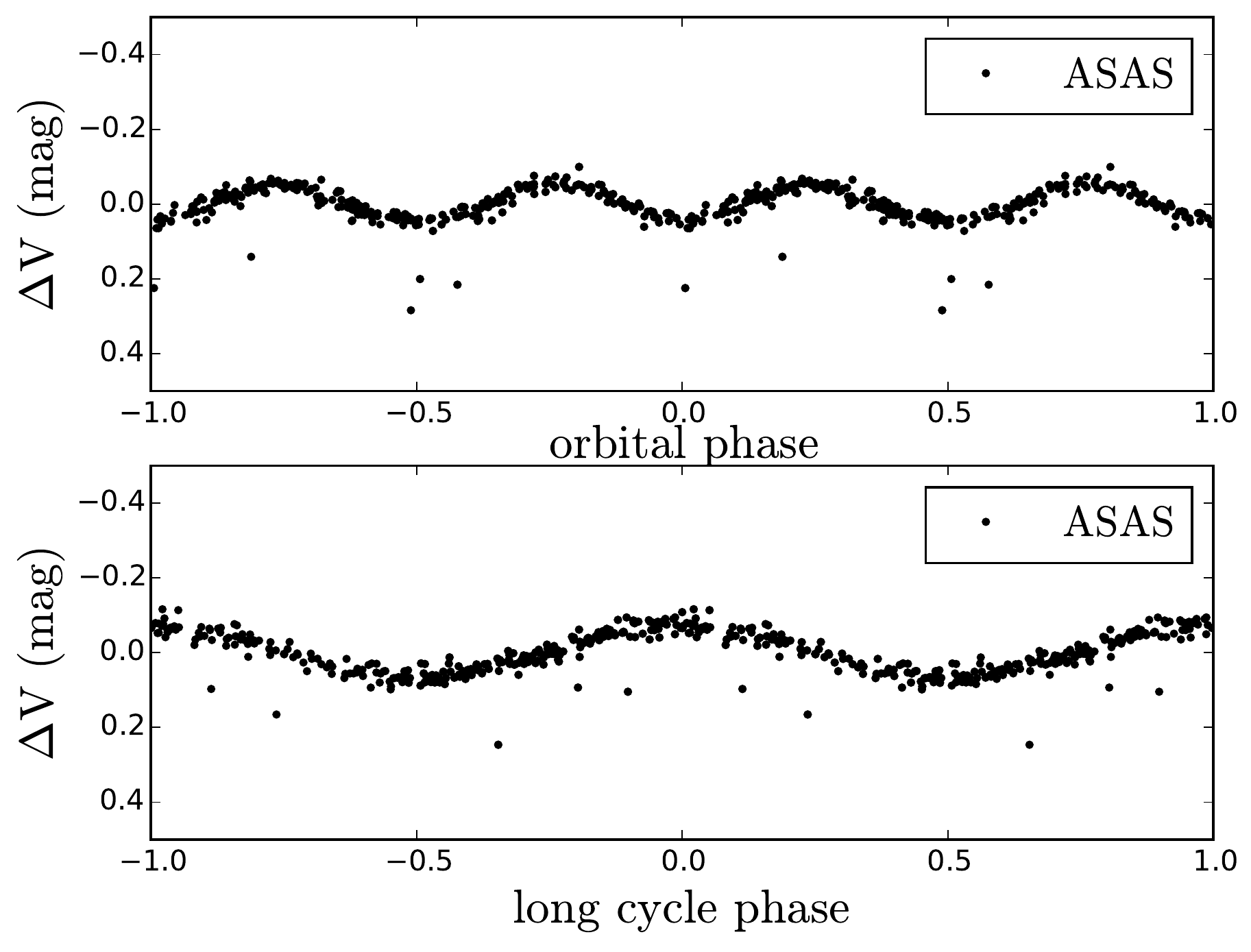}
	\end{center}
	\caption{ASAS $V$-band phased light curves after disentangling. Orbital (up) and long cycle (down) phases are shown. Phases were calculated according to times of light curve minimum and maxima respectively, as given by equations (\ref{eq: eq. 1}) and (\ref{eq: eq. 2}).}
	\label{fig:Fig. 1}
\end{figure}

Once obtained both periods, we disentangled the light curves using a code specially designed for this purpose by late Zbigniew Ko\l{}aczkowski. 
Briefly, the code adjusts a Fourier series consisting of the fundamental frequency plus their harmonics  to the orbital signal previous computed. Later, it removes the signal from the original time series allowing us to obtain the long periodicity and the orbital one in two separated light curves (Fig. \ref{fig:Fig. 1}). This code is described in more detail by \citet{2012MNRAS.421..862M}. The process reveals an orbital modulation typical of an ellipsoidal DPV  and a longer cycle characterized by a quasi-sinusoidal variability as usual in others double periodic variables. Both cycles are of comparable amplitude, contrary to the observed in some eclipsing DPVs like V495 Cen, where the orbital variability dominates \citep{2018MNRAS.476.3039R}. We have determined the following ephemeris for the light curves and these will be used for the analysis in the rest of the paper:

\begin{equation}
	\mathrm{HJD_{min,orbital}}= 2453096.54630 + 6.701(1) \times E,
	\label{eq: eq. 1}
\end{equation}
\begin{equation}
	\mathrm{HJD_{max,long}}= 2452942.84607 + 191(2)\times E,
	\label{eq: eq. 2}
\end{equation}

\noindent
wherein HJD corresponds to Heliocentric Julian Date labeled with their respective sub-indexes, i.e. for the  minimum orbital cycle (min, orbital) and for the maximum long cycle (max, long), while $E$ is an entire number reflecting the number of elapsed cycles. As we will show later, the main eclipse occurs when the more massive and hotter star is eclipsed. We will call primary star to this star, while the less massive and coldest is the secondary star. Since we find evidence of mass transfer from the cold to the hot star, we also name them donor and gainer, respectively.

\section{Spectroscopic observations}
\label{Sec: Sec. 3}

We have collected a series of 138 spectra uniformly distributed on orbital phase, of high/medium spectral resolution with different spectrographs during the years 2008 and 2015. We started the year 2008 using the CORALIE\footnote{\url{https://www.eso.org/public/chile/teles-instr/lasilla/swiss/coralie/}} spectrograph at the ESO La Silla Observatory, obtaining 37 spectra of signal to noise ratio $(\mathrm{SNR}) \approx 58$  with a resolving power $\mathrm{R} \sim 60000$ (Table \ref{Tab: Tab. 2}). Further, during  the years 2009, 2013 and part of the year 2015, we used the ECHELLE\footnote{\url{http://www.astrossp.unam.mx/oanspm/}} spectrograph in the San Pedro Mártir Observatory obtaining 49 spectra of $\mathrm{SNR} \approx 124$ with $\mathrm{R} \sim 18000$ (Table \ref{Tab: Tab. 3}).  In addition, between years 2012 to 2013 we used CHIRON\footnote{\url{http://www.ctio.noao.edu/noao/node/847}} spectrograph in Cerro Tololo Interamerican Observatory obtaining a sample of 52 spectra of $\mathrm{SNR} \approx 47$ (slicer mode) with $\mathrm{R} \sim 80000$ (Table \ref{Tab: Tab. 3}). The covered spectral region were 3875-6895 \AA{} (CORALIE), 3520-7605 \AA{} (ECHELLE) and 4580-8780 \AA{} (CHIRON); this provided us a broad spectral range for studying the variability of the spectral features of the stellar components. 

The correction of every spectra with flat, bias, and the wavelength calibration were done with \texttt{IRAF}, and the spectra were normalized to the continuum. The velocities are expressed in the heliocentric rest of frame. The signal to noise ratio ($S/N$) of every spectrum was computed  over a region of the continuum depleted of spectral lines. We have not sky subtracted neither flux calibrated our spectra, but this does not affect the strength of measurements neither the radial velocities (RVs). Our observations cover $ 100 \% $ of the orbital cycle, while for the long-term periodicity the coverage is about $ 90\% $.

\begin{table}[h!]
	\caption{Summary of spectroscopic observation using the CORALIE spectrograph mounted on EULER Telescope. The HJD at mid-exposure for the first spectrum series is given, the orbital ($\phi_\mathrm{o}$) long cycle ($\phi_\mathrm{l}$) phases were calculated using the eq.\,\ref{eq: eq. 1} and eq.\,\ref{eq: eq. 2} respectively. These spectra have a spectral resolution of $\mathrm{R \sim 60000}$.}
	\normalsize
	\begin{center}
		\resizebox{0.5\textwidth}{7cm}{		
			\begin{tabular}{cccccc}
				\hline
				\noalign
				{\smallskip}
				\textrm {UT-date} & \textrm {exptime (s)}  &\textrm {HJD} &\textrm {$\phi_\mathrm{o}$}   & \textrm {$\phi_\mathrm{l}$} &  $S/N$\\
				\hline
				\hline
				\noalign{\smallskip}
				\textrm{2008-04-05} & \textrm{ 600} & \textrm{2454562.31964120} & \textrm{0.7395} & \textrm{0.4700} & \textrm{46.14} \\
				\textrm{2008-04-05} & \textrm{ 600} & \textrm{2454562.32746528} & \textrm{0.7407} & \textrm{0.4700} & \textrm{49.06} \\
				\textrm{2008-04-05} & \textrm{ 900} & \textrm{2454562.33877315} & \textrm{0.7423} & \textrm{0.4701} & \textrm{62.89} \\
				\textrm{2008-04-05} & \textrm{ 900} & \textrm{2454562.35005787} & \textrm{0.7440} & \textrm{0.4701} & \textrm{58.76} \\
				\textrm{2008-04-06} & \textrm{1200} & \textrm{2454563.32796296} & \textrm{0.8900} & \textrm{0.4752} & \textrm{62.15} \\
				\textrm{2008-04-06} & \textrm{1200} & \textrm{2454563.34273148} & \textrm{0.8922} & \textrm{0.4753} & \textrm{68.61} \\
				\textrm{2008-04-06} & \textrm{1200} & \textrm{2454563.35752315} & \textrm{0.8944} & \textrm{0.4754} & \textrm{69.80} \\
				\textrm{2008-04-07} & \textrm{ 900} & \textrm{2454564.31726852} & \textrm{0.0376} & \textrm{0.4804} & \textrm{51.70} \\
				\textrm{2008-04-07} & \textrm{ 900} & \textrm{2454564.32856481} & \textrm{0.0393} & \textrm{0.4805} & \textrm{53.90} \\
				\textrm{2008-04-07} & \textrm{ 900} & \textrm{2454564.34033565} & \textrm{0.0410} & \textrm{0.4805} & \textrm{51.28} \\
				\textrm{2008-04-08} & \textrm{ 900} & \textrm{2454565.32734954} & \textrm{0.1883} & \textrm{0.4857} & \textrm{60.72} \\
				\textrm{2008-04-08} & \textrm{ 900} & \textrm{2454565.33864583} & \textrm{0.1900} & \textrm{0.4858} & \textrm{65.20} \\
				\textrm{2008-04-08} & \textrm{ 900} & \textrm{2454565.34995370} & \textrm{0.1917} & \textrm{0.4858} & \textrm{64.36} \\
				\textrm{2008-04-15} & \textrm{ 900} & \textrm{2454937.46894670} & \textrm{0.7236} & \textrm{0.4320} & \textrm{61.03} \\
				\textrm{2008-04-15} & \textrm{ 900} & \textrm{2454937.48075231} & \textrm{0.7253} & \textrm{0.4321} & \textrm{68.15} \\
				\textrm{2008-04-15} & \textrm{ 900} & \textrm{2454937.49209491} & \textrm{0.7270} & \textrm{0.4321} & \textrm{67.71} \\
				\textrm{2008-04-16} & \textrm{ 900} & \textrm{2454937.50343750} & \textrm{0.7287} & \textrm{0.4322} & \textrm{73.32} \\
				\textrm{2008-04-16} & \textrm{1200} & \textrm{2454938.47243056} & \textrm{0.8733} & \textrm{0.4373} & \textrm{71.75} \\
				\textrm{2008-04-16} & \textrm{1200} & \textrm{2454938.48769676} & \textrm{0.8756} & \textrm{0.4373} & \textrm{71.36} \\
				\textrm{2008-04-17} & \textrm{1200} & \textrm{2454938.50247685} & \textrm{0.8778} & \textrm{0.4374} & \textrm{65.43} \\
				\textrm{2008-04-17} & \textrm{1200} & \textrm{2454938.51725694} & \textrm{0.8800} & \textrm{0.4375} & \textrm{58.55} \\
				\textrm{2008-04-17} & \textrm{1200} & \textrm{2454938.53203704} & \textrm{0.8822} & \textrm{0.4376} & \textrm{59.53} \\
				\textrm{2008-05-16} & \textrm{ 900} & \textrm{2454968.45359954} & \textrm{0.3475} & \textrm{0.5941} & \textrm{53.94} \\
				\textrm{2008-05-16} & \textrm{ 900} & \textrm{2454968.46481481} & \textrm{0.3491} & \textrm{0.5941} & \textrm{56.56} \\
				\textrm{2008-05-22} & \textrm{ 900} & \textrm{2454609.32633102} & \textrm{0.7544} & \textrm{0.7158} & \textrm{47.55} \\
				\textrm{2008-05-22} & \textrm{ 900} & \textrm{2454609.33790509} & \textrm{0.7561} & \textrm{0.7159} & \textrm{43.55} \\
				\textrm{2008-05-23} & \textrm{ 900} & \textrm{2454610.29773148} & \textrm{0.8993} & \textrm{0.7209} & \textrm{56.35} \\
				\textrm{2008-05-23} & \textrm{ 900} & \textrm{2454610.30902778} & \textrm{0.9010} & \textrm{0.7209} & \textrm{49.34} \\
				\textrm{2008-05-23} & \textrm{ 900} & \textrm{2454610.32034722} & \textrm{0.9027} & \textrm{0.7210} & \textrm{42.64} \\
				\textrm{2008-10-02} & \textrm{ 999} & \textrm{2454741.67127315} & \textrm{0.5044} & \textrm{0.4080} & \textrm{57.29} \\
				\textrm{2008-10-02} & \textrm{ 999} & \textrm{2454741.68372685} & \textrm{0.5063} & \textrm{0.4080} & \textrm{48.16} \\
				\textrm{2008-10-03} & \textrm{1200} & \textrm{2454742.68658565} & \textrm{0.6559} & \textrm{0.4133} & \textrm{32.20} \\
				\textrm{2008-10-04} & \textrm{1200} & \textrm{2454743.66797454} & \textrm{0.8024} & \textrm{0.4184} & \textrm{61.84} \\
				\textrm{2008-10-04} & \textrm{1200} & \textrm{2454743.68747685} & \textrm{0.8053} & \textrm{0.4185} & \textrm{54.12} \\
				\textrm{2008-12-19} & \textrm{ 900} & \textrm{2454819.69328714} & \textrm{0.1477} & \textrm{0.8160} & \textrm{62.10} \\
				\textrm{2008-12-20} & \textrm{ 900} & \textrm{2454820.68160635} & \textrm{0.2952} & \textrm{0.8212} & \textrm{63.67} \\
				\textrm{2008-12-22} & \textrm{ 900} & \textrm{2454822.69305410} & \textrm{0.5954} & \textrm{0.8317} & \textrm{64.95} \\
				\hline
			\end{tabular}
		}
	\end{center}
	\label{Tab: Tab. 2}
\end{table}

\begin{table}[h!]
	\caption{Summary of spectroscopic observation using the ECHELLE spectrograph mounted on the San Pedro Mártir Telescope. The HJD at mid-exposure for the first spectrum series is given, the orbital ($\phi_\mathrm{o}$) and long cycle ($\phi_\mathrm{l}$) phases were calculated using eq.\,\ref{eq: eq. 1} and eq.\,\ref{eq: eq. 2} respectively with a spectral resolution of $\mathrm{R \sim 18000}$. }
	\normalsize
	\begin{center}
		\resizebox{0.5\textwidth}{9cm}{		
			\begin{tabular}{cccccc}
				\hline
				\noalign
				{\smallskip}
				\textrm {UT-date} & \textrm {exptime (s)}  &\textrm {HJD} &\textrm {$\phi_\mathrm{o}$}   & \textrm {$\phi_\mathrm{l}$} &  $S/N$\\
				\hline
				\hline
				\noalign{\smallskip}
				\textrm{2009-12-06} & \textrm{1200} & \textrm{2455171.91678429} & \textrm{0.7106} & \textrm{0.6582} & \textrm{143.05} \\
				\textrm{2009-12-06} & \textrm{1200} & \textrm{2455171.93134509} & \textrm{0.7127} & \textrm{0.6583} & \textrm{178.73} \\
				\textrm{2009-12-06} & \textrm{1200} & \textrm{2455171.94591749} & \textrm{0.7149} & \textrm{0.6584} & \textrm{131.58} \\
				\textrm{2009-12-07} & \textrm{1200} & \textrm{2455172.98771418} & \textrm{0.8704} & \textrm{0.6638} & \textrm{ 30.89} \\
				\textrm{2009-12-07} & \textrm{1200} & \textrm{2455173.00228659} & \textrm{0.8726} & \textrm{0.6639} & \textrm{ 52.55} \\
				\textrm{2009-11-04} & \textrm{1200} & \textrm{2455140.04944751} & \textrm{0.9550} & \textrm{0.4915} & \textrm{ 50.68} \\
				\textrm{2009-11-05} & \textrm{1200} & \textrm{2455141.04128960} & \textrm{0.1030} & \textrm{0.4967} & \textrm{135.79} \\
				\textrm{2009-11-07} & \textrm{1200} & \textrm{2455142.98124607} & \textrm{0.3925} & \textrm{0.5069} & \textrm{157.92} \\
				\textrm{2009-11-08} & \textrm{1200} & \textrm{2455143.94206029} & \textrm{0.5359} & \textrm{0.5119} & \textrm{142.01} \\
				\textrm{2009-11-09} & \textrm{ 900} & \textrm{2455144.87802708} & \textrm{0.6755} & \textrm{0.5168} & \textrm{ 89.46} \\
				\textrm{2009-11-09} & \textrm{ 900} & \textrm{2455144.88912728} & \textrm{0.6772} & \textrm{0.5168} & \textrm{116.88} \\
				\textrm{2009-11-09} & \textrm{ 900} & \textrm{2455144.90022751} & \textrm{0.6788} & \textrm{0.5169} & \textrm{135.28} \\
				\textrm{2009-11-10} & \textrm{ 900} & \textrm{2455146.01491087} & \textrm{0.8452} & \textrm{0.5227} & \textrm{129.14} \\
				\textrm{2009-11-10} & \textrm{ 900} & \textrm{2455146.02601127} & \textrm{0.8468} & \textrm{0.5228} & \textrm{170.71} \\
				\textrm{2009-11-10} & \textrm{ 900} & \textrm{2455146.03710007} & \textrm{0.8485} & \textrm{0.5228} & \textrm{134.24} \\
				\textrm{2011-01-25} & \textrm{ 900} & \textrm{2455586.76571973} & \textrm{0.6191} & \textrm{0.8279} & \textrm{122.56} \\
				\textrm{2011-01-26} & \textrm{ 900} & \textrm{2455587.76936629} & \textrm{0.7689} & \textrm{0.8331} & \textrm{ 88.93} \\
				\textrm{2011-01-27} & \textrm{ 900} & \textrm{2455588.69713235} & \textrm{0.9073} & \textrm{0.8380} & \textrm{101.30} \\
				\textrm{2011-01-28} & \textrm{ 900} & \textrm{2455589.70166633} & \textrm{0.0572} & \textrm{0.8432} & \textrm{103.31} \\
				\textrm{2012-02-11} & \textrm{1200} & \textrm{2455968.82382656} & \textrm{0.6342} & \textrm{0.8261} & \textrm{186.58} \\
				\textrm{2012-02-12} & \textrm{1200} & \textrm{2455969.88029338} & \textrm{0.7918} & \textrm{0.8316} & \textrm{152.52} \\
				\textrm{2012-02-13} & \textrm{1200} & \textrm{2455970.75980939} & \textrm{0.9231} & \textrm{0.8362} & \textrm{175.94} \\
				\textrm{2012-02-14} & \textrm{1200} & \textrm{2455971.78222342} & \textrm{0.0756} & \textrm{0.8416} & \textrm{207.17} \\
				\textrm{2012-02-15} & \textrm{1200} & \textrm{2455972.87665882} & \textrm{0.2390} & \textrm{0.8473} & \textrm{ 13.43} \\
				\textrm{2012-02-18} & \textrm{1200} & \textrm{2455975.70836929} & \textrm{0.6616} & \textrm{0.8621} & \textrm{120.36} \\
				\textrm{2012-02-18} & \textrm{1200} & \textrm{2455975.79925421} & \textrm{0.6751} & \textrm{0.8626} & \textrm{154.01} \\
				\textrm{2012-02-18} & \textrm{1200} & \textrm{2455975.86411098} & \textrm{0.6848} & \textrm{0.8629} & \textrm{124.37} \\
				\textrm{2012-02-19} & \textrm{1200} & \textrm{2455976.61805307} & \textrm{0.7973} & \textrm{0.8668} & \textrm{182.46} \\
				\textrm{2012-02-19} & \textrm{1200} & \textrm{2455976.78931463} & \textrm{0.8229} & \textrm{0.8677} & \textrm{129.46} \\
				\textrm{2012-02-20} & \textrm{1200} & \textrm{2455977.64297014} & \textrm{0.9503} & \textrm{0.8722} & \textrm{174.55} \\
				\textrm{2012-02-20} & \textrm{1200} & \textrm{2455977.73496695} & \textrm{0.9640} & \textrm{0.8727} & \textrm{144.43} \\
				\textrm{2012-02-21} & \textrm{1200} & \textrm{2455978.75980151} & \textrm{0.1169} & \textrm{0.8780} & \textrm{ 73.49} \\
				\textrm{2012-02-21} & \textrm{1200} & \textrm{2455978.77386302} & \textrm{0.1190} & \textrm{0.8781} & \textrm{ 51.67} \\
				\textrm{2012-02-22} & \textrm{1200} & \textrm{2455979.70671364} & \textrm{0.2582} & \textrm{0.8830} & \textrm{195.94} \\
				\textrm{2012-02-22} & \textrm{1200} & \textrm{2455979.72076366} & \textrm{0.2603} & \textrm{0.8831} & \textrm{175.80} \\
				\textrm{2012-02-23} & \textrm{1200} & \textrm{2455980.70032159} & \textrm{0.4065} & \textrm{0.8882} & \textrm{138.41} \\
				\textrm{2012-11-13} & \textrm{1200} & \textrm{2456245.02839128} & \textrm{0.8526} & \textrm{0.2706} & \textrm{215.02} \\
				\textrm{2012-11-14} & \textrm{1200} & \textrm{2456245.91058693} & \textrm{0.9842} & \textrm{0.2753} & \textrm{182.54} \\
				\textrm{2012-11-18} & \textrm{1200} & \textrm{2456250.00101000} & \textrm{0.5946} & \textrm{0.2967} & \textrm{227.72} \\
				\textrm{2012-11-19} & \textrm{1200} & \textrm{2456250.87730000} & \textrm{0.7254} & \textrm{0.3012} & \textrm{119.71} \\
				\textrm{2013-01-28} & \textrm{1200} & \textrm{2456320.72825435} & \textrm{0.1494} & \textrm{0.6666} & \textrm{183.37} \\
				\textrm{2013-01-28} & \textrm{1200} & \textrm{2456320.74231642} & \textrm{0.1515} & \textrm{0.6666} & \textrm{158.04} \\
				\textrm{2013-01-28} & \textrm{1200} & \textrm{2456320.75639001} & \textrm{0.1536} & \textrm{0.6667} & \textrm{144.43} \\
				\textrm{2013-01-30} & \textrm{1200} & \textrm{2456322.73900591} & \textrm{0.4494} & \textrm{0.6771} & \textrm{130.41} \\
				\textrm{2013-01-30} & \textrm{1200} & \textrm{2456322.75306786} & \textrm{0.4515} & \textrm{0.6772} & \textrm{132.03} \\
				\textrm{2013-01-30} & \textrm{1200} & \textrm{2456322.76714142} & \textrm{0.4536} & \textrm{0.6772} & \textrm{139.45} \\
				\textrm{2015-01-13} & \textrm{1200} & \textrm{2457035.95402000} & \textrm{0.8836} & \textrm{0.4072} & \textrm{215.92} \\
				\textrm{2015-01-13} & \textrm{1200} & \textrm{2457035.96816000} & \textrm{0.8857} & \textrm{0.4073} & \textrm{159.34} \\
				\textrm{2015-01-13} & \textrm{1200} & \textrm{2457035.98227000} & \textrm{0.8878} & \textrm{0.4074} & \textrm{144.10} \\
				\hline
			\end{tabular}
		}
	\end{center}
	\label{Tab: Tab. 3}
\end{table}

\begin{table}[h!]
	\caption{Summary of spectroscopic observation using the CHIRON spectrograph. The HJD at mid-exposure for the first spectrum series is given, the orbital ($\phi_\mathrm{o}$) and long cycle ($\phi_\mathrm{l}$) phases were calculated using the eq. \ref{eq: eq. 1} and eq. \ref{eq: eq. 2} respectively with a spectral resolution of $\mathrm{R \sim 80000}$ (Slicer mode).}
	\normalsize
	\begin{center}
		\resizebox{0.5\textwidth}{9.5cm}{		
			\begin{tabular}{cccccc}
				\hline
				\noalign
				{\smallskip}
				\textrm {UT-date} & \textrm {exptime (s)}  &\textrm {HJD} &\textrm {$\phi_\mathrm{o}$}   & \textrm {$\phi_\mathrm{l}$} &  $S/N$\\
				\hline
				\hline
				\noalign{\smallskip}
				\textrm{2012-11-05} &\textrm{1200} &  \textrm{2456236.80526037} & \textrm{0.6254} & \textrm{0.2276} & \textrm{64.30} \\
				\textrm{2012-11-08} &\textrm{1200} &  \textrm{2456239.79422590} & \textrm{0.0715} & \textrm{0.2433} & \textrm{58.38} \\
				\textrm{2012-11-08} &\textrm{1200} &  \textrm{2456239.80832071} & \textrm{0.0736} & \textrm{0.2433} & \textrm{62.83} \\
				\textrm{2012-11-08} &\textrm{1200} &  \textrm{2456239.82241444} & \textrm{0.0757} & \textrm{0.2434} & \textrm{61.62} \\
				\textrm{2012-11-11} &\textrm{1200} &  \textrm{2456242.82871200} & \textrm{0.5243} & \textrm{0.2591} & \textrm{66.10} \\
				\textrm{2012-11-11} &\textrm{1200} &  \textrm{2456242.84280691} & \textrm{0.5264} & \textrm{0.2592} & \textrm{63.63} \\
				\textrm{2012-11-11} &\textrm{1200} &  \textrm{2456242.85690172} & \textrm{0.5285} & \textrm{0.2593} & \textrm{60.01} \\
				\textrm{2012-11-14} &\textrm{1200} &  \textrm{2456245.78951280} & \textrm{0.9662} & \textrm{0.2746} & \textrm{73.99} \\
				\textrm{2012-11-14} &\textrm{1200} &  \textrm{2456245.80360880} & \textrm{0.9683} & \textrm{0.2747} & \textrm{69.66} \\
				\textrm{2012-11-14} &\textrm{1200} &  \textrm{2456245.81770132} & \textrm{0.9704} & \textrm{0.2748} & \textrm{47.09} \\
				\textrm{2012-11-20} &\textrm{1200} &  \textrm{2456251.73843209} & \textrm{0.8539} & \textrm{0.3057} & \textrm{73.50} \\
				\textrm{2012-11-20} &\textrm{1200} &  \textrm{2456251.75252803} & \textrm{0.8560} & \textrm{0.3058} & \textrm{68.08} \\
				\textrm{2012-11-20} &\textrm{1200} &  \textrm{2456251.76662151} & \textrm{0.8581} & \textrm{0.3059} & \textrm{73.05} \\
				\textrm{2012-11-25} &\textrm{1200} &  \textrm{2456256.83286117} & \textrm{0.6142} & \textrm{0.3324} & \textrm{62.18} \\
				\textrm{2012-11-28} &\textrm{1200} &  \textrm{2456259.70042938} & \textrm{0.0421} & \textrm{0.3474} & \textrm{54.80} \\
				\textrm{2012-11-28} &\textrm{1200} &  \textrm{2456259.71452406} & \textrm{0.0442} & \textrm{0.3475} & \textrm{50.92} \\
				\textrm{2012-11-28} &\textrm{1200} &  \textrm{2456259.72861750} & \textrm{0.0463} & \textrm{0.3475} & \textrm{48.31} \\
				\textrm{2012-12-04} &\textrm{1200} &  \textrm{2456265.78505374} & \textrm{0.9501} & \textrm{0.3792} & \textrm{63.94} \\
				\textrm{2012-12-04} &\textrm{1200} &  \textrm{2456265.79914814} & \textrm{0.9522} & \textrm{0.3793} & \textrm{57.33} \\
				\textrm{2012-12-04} &\textrm{1200} &  \textrm{2456265.81324131} & \textrm{0.9543} & \textrm{0.3794} & \textrm{54.12} \\
				\textrm{2013-11-04} &\textrm{1200} &  \textrm{2456600.84049212} & \textrm{0.9509} & \textrm{0.1316} & \textrm{41.22} \\
				\textrm{2013-11-04} &\textrm{1200} &  \textrm{2456600.85227210} & \textrm{0.9527} & \textrm{0.1316} & \textrm{29.08} \\
				\textrm{2013-11-04} &\textrm{1200} &  \textrm{2456600.86405207} & \textrm{0.9544} & \textrm{0.1317} & \textrm{26.02} \\
				\textrm{2013-11-06} &\textrm{1200} &  \textrm{2456602.82314142} & \textrm{0.2468} & \textrm{0.1419} & \textrm{53.63} \\
				\textrm{2013-11-06} &\textrm{1200} &  \textrm{2456602.83723646} & \textrm{0.2489} & \textrm{0.1420} & \textrm{49.99} \\
				\textrm{2013-11-06} &\textrm{1200} &  \textrm{2456602.85133027} & \textrm{0.2510} & \textrm{0.1421} & \textrm{49.40} \\
				\textrm{2013-11-08} &\textrm{1200} &  \textrm{2456604.78689240} & \textrm{0.5399} & \textrm{0.1522} & \textrm{56.14} \\
				\textrm{2013-11-08} &\textrm{1200} &  \textrm{2456604.80098707} & \textrm{0.5420} & \textrm{0.1523} & \textrm{57.10} \\
				\textrm{2013-11-08} &\textrm{1200} &  \textrm{2456604.81508074} & \textrm{0.5441} & \textrm{0.1524} & \textrm{57.05} \\
				\textrm{2013-11-12} &\textrm{1200} &  \textrm{2456608.84935936} & \textrm{0.1461} & \textrm{0.1735} & \textrm{49.36} \\
				\textrm{2013-11-12} &\textrm{1200} &  \textrm{2456608.86345529} & \textrm{0.1482} & \textrm{0.1735} & \textrm{49.10} \\
				\textrm{2013-11-12} &\textrm{1200} &  \textrm{2456608.87754884} & \textrm{0.1503} & \textrm{0.1736} & \textrm{51.28} \\
				\textrm{2013-11-22} &\textrm{1200} &  \textrm{2456618.78744321} & \textrm{0.6292} & \textrm{0.2254} & \textrm{49.51} \\
				\textrm{2013-11-22} &\textrm{1200} &  \textrm{2456618.80153801} & \textrm{0.6313} & \textrm{0.2255} & \textrm{38.32} \\
				\textrm{2013-11-22} &\textrm{1200} &  \textrm{2456618.81563157} & \textrm{0.6334} & \textrm{0.2256} & \textrm{46.79} \\
				\textrm{2013-11-17} &\textrm{1200} &  \textrm{2456613.74164809} & \textrm{0.8762} & \textrm{0.1990} & \textrm{50.55} \\
				\textrm{2013-11-17} &\textrm{1200} &  \textrm{2456613.75574294} & \textrm{0.8783} & \textrm{0.1991} & \textrm{50.10} \\
				\textrm{2013-11-17} &\textrm{1200} &  \textrm{2456613.76983652} & \textrm{0.8804} & \textrm{0.1992} & \textrm{47.54} \\
				\textrm{2013-11-18} &\textrm{1200} &  \textrm{2456614.79142646} & \textrm{0.0328} & \textrm{0.2045} & \textrm{48.91} \\
				\textrm{2013-11-18} &\textrm{1200} &  \textrm{2456614.80552017} & \textrm{0.0350} & \textrm{0.2046} & \textrm{53.69} \\
				\textrm{2013-11-18} &\textrm{1200} &  \textrm{2456614.81961616} & \textrm{0.0371} & \textrm{0.2047} & \textrm{49.32} \\
				\textrm{2013-11-20} &\textrm{1200} &  \textrm{2456616.80744009} & \textrm{0.3337} & \textrm{0.2151} & \textrm{54.65} \\
				\textrm{2013-11-20} &\textrm{1200} &  \textrm{2456616.82153479} & \textrm{0.3358} & \textrm{0.2152} & \textrm{49.41} \\
				\textrm{2013-11-20} &\textrm{1200} &  \textrm{2456616.83562947} & \textrm{0.3379} & \textrm{0.2152} & \textrm{50.11} \\
				\textrm{2013-11-24} &\textrm{1200} &  \textrm{2456620.71242225} & \textrm{0.9164} & \textrm{0.2355} & \textrm{23.26} \\
				\textrm{2013-11-24} &\textrm{1200} &  \textrm{2456620.72651699} & \textrm{0.9186} & \textrm{0.2356} & \textrm{27.04} \\
				\textrm{2013-11-24} &\textrm{1200} &  \textrm{2456620.74061160} & \textrm{0.9207} & \textrm{0.2356} & \textrm{24.93} \\
				\textrm{2013-11-26} &\textrm{1200} &  \textrm{2456622.78521955} & \textrm{0.2258} & \textrm{0.2463} & \textrm{45.55} \\
				\textrm{2013-11-26} &\textrm{1200} &  \textrm{2456622.81340771} & \textrm{0.2300} & \textrm{0.2465} & \textrm{47.72} \\
				\textrm{2013-12-16} &\textrm{1200} &  \textrm{2456642.74995625} & \textrm{0.2051} & \textrm{0.3508} & \textrm{22.40} \\
				\textrm{2013-12-16} &\textrm{1200} &  \textrm{2456642.76405042} & \textrm{0.2072} & \textrm{0.3508} & \textrm{25.21} \\
				\textrm{2013-12-16} &\textrm{1200} &  \textrm{2456642.77814343} & \textrm{0.2093} & \textrm{0.3509} & \textrm{26.70} \\
				\hline
			\end{tabular}
		}
	\end{center}
	\label{Tab: Tab. 4}
\end{table}


\section{Spectroscopic analysis}
\label{Sec: Sec. 4}

In this section we provide the analysis of our spectra, revealing  constrains on the system parameters and evidencing the presence of circumstellar matter in the system.

\subsection{Radial velocities for the donor}
\label{Subsec: Subsec. 4.1}

Close binary star orbits can be characterized through the study of the radial velocities of their stellar components. We selected some characteristic lines that represent the movement of both components. The radial velocity of the donor was obtained by Gaussian fit  on a set of H$\alpha$ lines from spectra obtained with the CORALIE spectrograph. In addition, the velocities were corrected to the absolute heliocentric system and the orbital parameters for HD\,50526 were obtained using the genetic algorithm {\texttt{PIKAIA}}\footnote{ \url{http://www.hao.ucar.edu/modeling/pikaia/pikaia.php}} of public domain developed by \citet{1995ApJS..101..309C}. This code produces a series of theoretical velocities and find the best parameters through the minimization of the function $\chi2$, defined as:

\noindent
$\chi2(P_\mathrm{o},\tau,\omega,e,K_{2},\gamma)$
\begin{equation}
	= \frac{1}{\mathrm{N}-6}\sum_\mathrm{j=1}^\mathrm{N} \left(\frac{V_\mathrm{j}^\mathrm{obs}-V(t_\mathrm{j};P_\mathrm{o},\tau,\omega,e,K_{2},\gamma)}{\sigma_\mathrm{j}}  \right)^2,
	\label{eq: eq. 3}
\end{equation}

\noindent
where the parameter $\mathrm{N}$ represents the number of observations, $V_\mathrm{j}^\mathrm{obs}$ is the radial velocity observed in the data set and $V(t_\mathrm{j};P_\mathrm{o},\omega,e,K_{2},\gamma)$ is the radial velocity at the time $t_\mathrm{j}$. $P_\mathrm{o}$ is the orbital period, $\tau$ the time of passage per the periastron, $\omega$ the periastron longitude, $e$ the orbital eccentricity, $K_{2}$ the half-amplitude of the radial velocities of the donor, and finally $\gamma$ the velocity of the center of mass of the system. According to the equation 2.45 given by \citet{2001icbs.book.....H} the theoretical radial velocity is given by :

\begin{equation}
	V(t)=\gamma + K_{2}(\cos(\omega+\theta(t)) +e{\cos(\omega)}),
	\label{eq: eq. 4}
\end{equation}

\noindent
where the angular parameter called true anomaly $\theta$ is obtained solving the relationship between the true and eccentric anomalies through the following equation:

\begin{equation}
	\tan\left(\frac{\theta}{2}\right)= \sqrt{\frac{1+e}{1-e}}\tan\left(\frac{E}{2}\right),
	\label{eq: eq. 5}
\end{equation}

\noindent
wherein $E$ corresponds to the eccentric anomaly and it is given by the equation 2.35 of \citet{2001icbs.book.....H}:

\begin{equation}
	E-e\sin(E) = \frac{2\pi}{P_{\mathrm{o}}}(t-\tau),
	\label{eq: eq. 6}
\end{equation}

\begin{figure}
\begin{center}
\includegraphics[trim=0.0cm 0.0cm 0.0cm 0.0cm,clip,width=0.45\textwidth,angle=0]{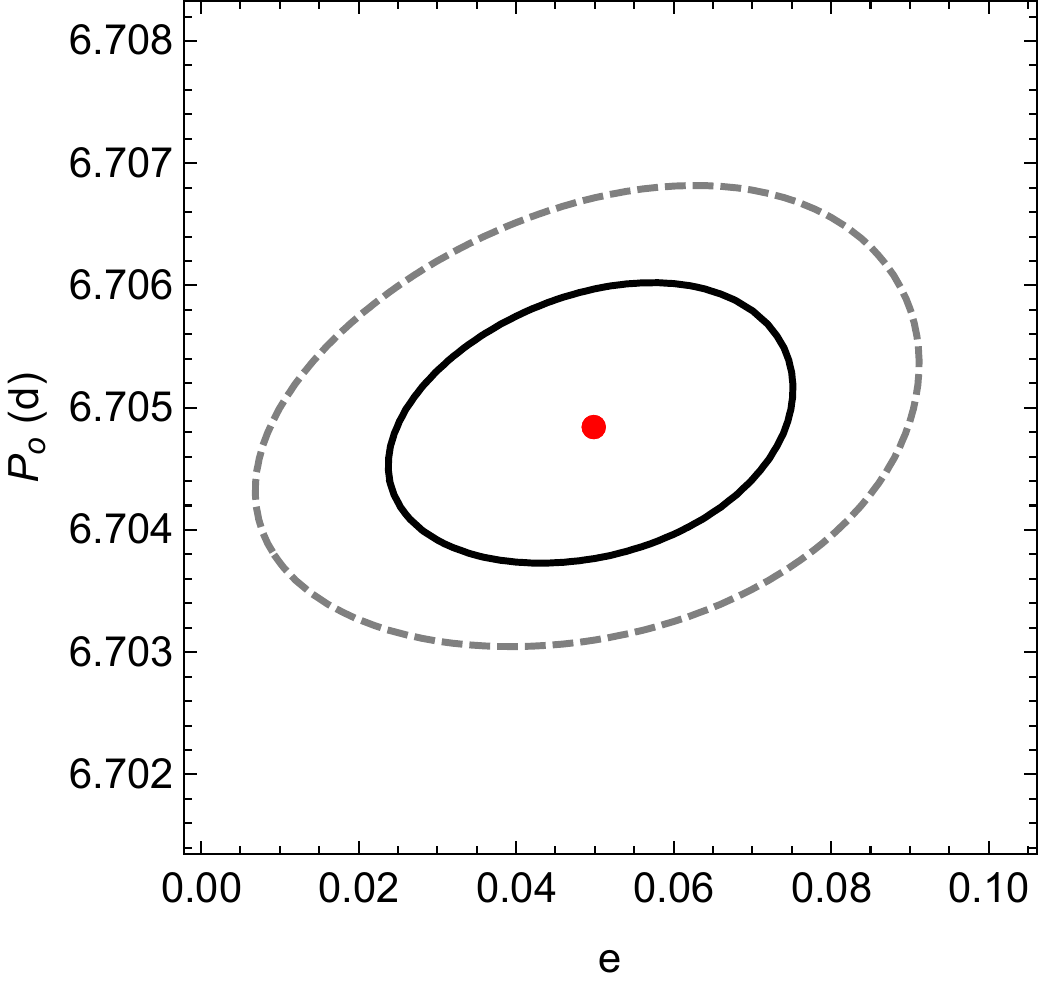}
\includegraphics[trim=0.0cm 0.0cm 0.0cm 0.0cm,clip,width=0.45\textwidth,angle=0]{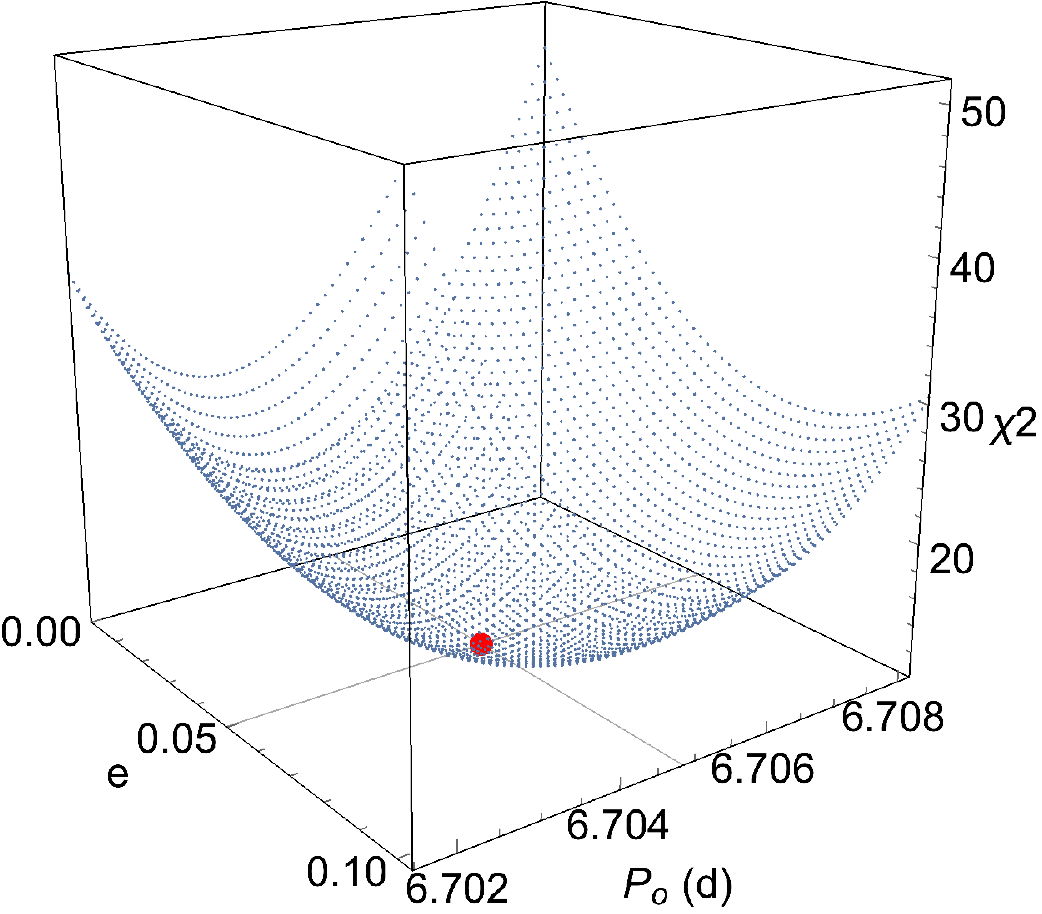}
\end{center}
\caption{(Up) The orbital period as a function of the eccentricity calculated by Monte Carlo simulations. The solid black line corresponds to $1\sigma$ isophote, the dashed gray line to $2\sigma$ isophote, and the red dot shows the minimum $\chi2$. (Down) The contour corresponds to the $\Delta \chi2$ of 2 degrees of freedom that includes 68.3\% of probability, i.e. $\Delta \chi2= 2.30$.}
	\label{fig:Fig. 2}
\end{figure}

\noindent
In order to solve the equation \ref{eq: eq. 5}, it is necessary to solve the equation \ref{eq: eq. 6}, but since it has not an analytic solution it must be solved numerically through an iterative method. After obtaining a solution for the true anomaly it is possible to obtain a solution for the equation \ref{eq: eq. 4} of the theoretical radial velocity $V(t_\mathrm{j};P_\mathrm{o},\omega,e,K_{2},\gamma)$. Through a Monte Carlo simulation we have estimated the error, perturbing the best fitting solution obtained with the \texttt{PIKAIA} code and computing a $\chi2$ for these perturbed solutions. As criterion for deciding when to accept an elliptical ($p<0.05$) or circular  ($p \geq 0.05$) orbit, we have implemented the statistical test $p_{1}$ of \citet{1971AJ.....76..544L} :

\begin{equation}
	p_{1}= \left(\frac{\sum (o-c)^{2}_\mathrm{ecc}}{\sum (o-c)^{2}_\mathrm{circ}}\right)^\mathrm{(n-m)/2},
	\label{eq: eq. 7}
\end{equation}

\noindent
the subindex \emph{ecc} corresponds to the $\chi2$ obtained from the fit of an elliptical orbit, while \emph{circ} is the $\chi2$ from a circular orbit, \emph{n} is the total number of observational radial velocities and \emph{m} is the number of free parameters for the fit of an elliptical orbit. Finally, we obtained the orbital parameters with its respective errors that are given in Table \ref{Tab: Tab. 5} with a value $p_{1}=0.0015$ compatible with an elliptical orbit, as is indicated in the Fig. \ref{fig:Fig. 2}. In this figure the solid black and the dashed gray lines correspond to $1\sigma$ and $2\sigma$ around the best solution marked with a red dot, respectively. It has been pointed out that gas stream and circumstellar matter can distort spectroscopic features in semidetached interacting binaries, producing skewed RVs and artificial small eccentricities \citep{2005A&A...439..663L}. For a non interacting binary with the stellar and orbital parameters of HD 50526 dynamical tides should have circularized the orbit and synchronized the rotational periods \citep{1975A&A....41..329Z, 1977A&A....57..383Z}. This should imply that the observed small eccentricity might be spurious. In addition, we find a small $2\sigma$ difference between the spectroscopic and photometric period. However, due to the larger density of data and time baseline of the photometric time series we trust more in the period obtained from photometry.

\begin{figure}
\begin{center}
\includegraphics[trim=0.2cm 0.2cm 0.2cm 0.2cm,clip,width=0.45\textwidth,angle=00]{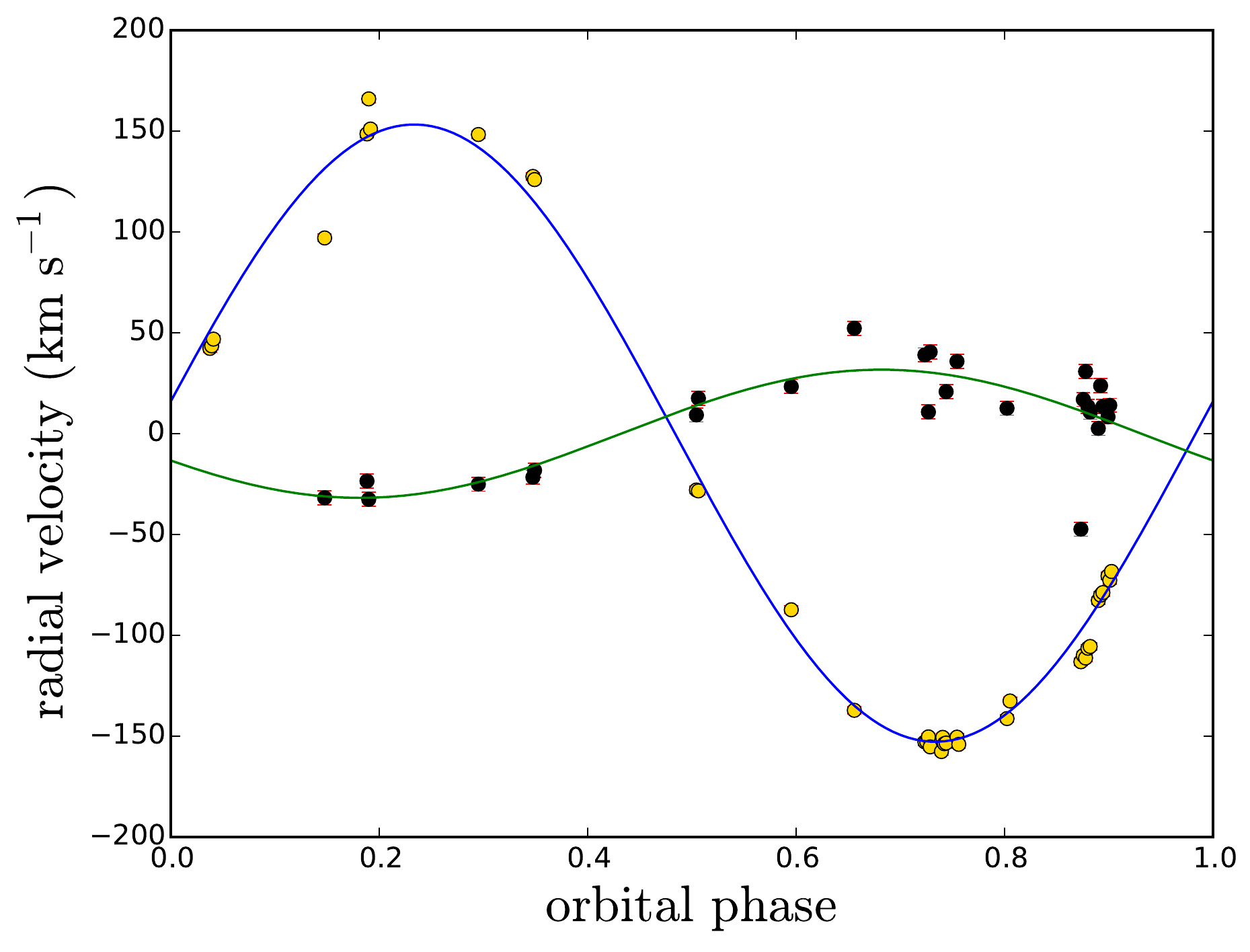}
\end{center}
\caption{Radial velocities of He\,I\,4471.477 and H$\alpha$  lines for the primary component (black dots) and for the secondary component (yellow dots) respectively. All measurements were through Gaussian fits. The best fitting functions are also shown.}
	\label{fig:Fig. 3}
\end{figure}

The radial velocities for the donor and gainer stars were measured through Gaussian fits in the CORALIE spectra, that turned to be the best spectra for this task.
These velocities are listed in Tables \ref{Tab: Tab. 6} and \ref{Tab: Tab. 7}. The radial velocities (RV) of the gainer star for He\,I\,4471.477 \AA{} was fitted with a sine function of amplitude $31.8 \pm 5.0 ~\mathrm{km\,s^{-1}}$ and zero point $0.0 \pm 3.4 ~\mathrm{km\,s^{-1}}$. The RV of the donor measured with the H$\alpha$ line, was fitted with an amplitude of $154.0 \pm 2.3 ~\mathrm{km\,s^{-1}}$ and a zero point $0.0 \pm 1.8 ~\mathrm{km\,s^{-1}}$. Both solutions assume a non-circular orbit and yield a mass ratio $q\equiv K_{1}/K_{2} = 0.206 \pm 0.033$ (Fig. \ref{fig:Fig. 3}).

\begin{table}{}
	\footnotesize
	\caption{Orbital elements for the donor of HD\,50526 obtained through minimization of $\chi2$ given  by equation (1). The value $\tau^{*}= \tau-2450000$ is given and the maximum and minimum are one isophote $1\sigma$.}
	\normalsize
	\[
	\begin{array}{l r r r}
	\hline
	\noalign{\smallskip}
	\mathrm {Parameter}   		& \mathrm{Best ~value}	& \mathrm{Low ~limit} 	& \mathrm{Upper ~limit}\\
	\hline
	\hline
	\noalign{\smallskip}
	P_\mathrm{o} ~(\mathrm{d})  & {6.7048}   			& {6.7034}  			& {6.7063}   	\\
	\tau^{*}              		& {566.392}				& {566.348}				& {566.436}  	\\
	\emph{e}              		& {0.04974}				& {0.02200}  			& {0.07600}  	\\
	\omega ~(\mathrm{rad})  	& {0.76636} 			& {0.72639}  			& {0.80639} 	\\
	K_{2} ~($\kms$)     		& {153.008}				& {149.158}				& {156.958}  	\\
	\gamma ~($\kms$)    		& {-5.16994} 			& {-7.93987} 			& {-2.33987}  	\\
	\hline
	\end{array}
	\]
	\label{Tab: Tab. 5}
\end{table}

\begin{table*}[h!]
	\caption{Radial velocities of the donor using the H$\alpha$ line in CORALIE spectra. Typical error is 1.9 km s$^{-1}$ and HJD' stands for HJD-2454000.}
	\normalsize
	\begin{center}
		\begin{tabular}{ccccccccc}
			\hline
			\noalign
			{\smallskip}
			\textrm {HJD'} & \textrm {$\phi_{o}$}  &\textrm{RV (km\,s$^{-1}$)} &	\textrm {HJD'} & \textrm {$\phi_{o}$}  &\textrm{RV (km\,s$^{-1}$)}& 	\textrm {HJD'} & \textrm {$\phi_{o}$}  &\textrm{RV (km\,s$^{-1}$)} \\
			\hline
			\hline
			\noalign{\smallskip}
			\textrm{564.31726852} & \textrm{0.0376} & \textrm{  42.3} &  \textrm{937.46894670} & \textrm{0.7236} & \textrm{-152.8} &\textrm{938.50247685} & \textrm{0.8778} & \textrm{-111.2}  \\  
			\textrm{564.32856481} & \textrm{0.0393} & \textrm{  43.6} &\textrm{937.48075231} & \textrm{0.7253} & \textrm{-152.6} & \textrm{938.51725694} & \textrm{0.8800} & \textrm{-106.3}  \\ 
			\textrm{564.34033565} & \textrm{0.0410} & \textrm{  46.9} &  \textrm{937.49209491} & \textrm{0.7270} & \textrm{-150.3} &\textrm{938.53203704} & \textrm{0.8822} & \textrm{-105.5} \\
			\textrm{819.69328714} & \textrm{0.1477} & \textrm{  97.1} &  \textrm{937.50343750} & \textrm{0.7287} & \textrm{-155.2} &	\textrm{563.32796296} & \textrm{0.8900} & \textrm{ -82.7} \\
			\textrm{565.32734954} & \textrm{0.1883} & \textrm{ 148.7} & \textrm{562.31964120} & \textrm{0.7395} & \textrm{-157.6} & 	\textrm{563.34273148} & \textrm{0.8922} & \textrm{ -80.0} \\
			\textrm{565.33864583} & \textrm{0.1900} & \textrm{ 166.0} &  \textrm{562.32746528} & \textrm{0.7407} & \textrm{-150.5} &\textrm{563.35752315} & \textrm{0.8944} & \textrm{ -78.7} \\
			\textrm{565.34995370} & \textrm{0.1917} & \textrm{ 151.1} & \textrm{562.33877315} & \textrm{0.7423} & \textrm{-153.7} & \textrm{610.29773148} & \textrm{0.8993} & \textrm{ -70.5} \\
			\textrm{820.68160635} & \textrm{0.2952} & \textrm{ 148.4} & \textrm{562.35005787} & \textrm{0.7440} & \textrm{-153.3} & 	\textrm{610.30902778} & \textrm{0.9010} & \textrm{ -72.7} \\
			\textrm{968.45359954} & \textrm{0.3475} & \textrm{ 127.6} & \textrm{609.32633102} & \textrm{0.7544} & \textrm{-150.4} & 	\textrm{610.32034722} & \textrm{0.9027} & \textrm{ -68.3} \\
			\textrm{968.46481481} & \textrm{0.3491} & \textrm{ 126.0} & \textrm{609.33790509} & \textrm{0.7561} & \textrm{-154.0} & & & \\
			\textrm{741.67127315} & \textrm{0.5044} & \textrm{ -27.9} & \textrm{743.66797454} & \textrm{0.8024} & \textrm{-141.2} & & & \\
			\textrm{741.68372685} & \textrm{0.5063} & \textrm{ -28.3} &  \textrm{743.68747685} & \textrm{0.8053} & \textrm{-132.5} & & & \\
			\textrm{822.69305410} & \textrm{0.5954} & \textrm{ -87.3} & \textrm{938.47243056} & \textrm{0.8733} & \textrm{-113.0} & & & \\
			\textrm{742.68658565} & \textrm{0.6559} & \textrm{-137.1} & \textrm{938.48769676} & \textrm{0.8756} & \textrm{-109.9} & & & \\
			\hline
		\end{tabular}
		
	\end{center}
	\label{Tab: Tab. 6}
\end{table*}

\begin{table*}[h!]
	\caption{Radial velocities of the gainer and their respective errors, using the He\,I\,4471.477 \AA{} line in CORALIE spectra. Typical error is 3.5 km s$^{-1}$ and HJD' stands for HJD-2454000.}
	\normalsize
	\begin{center}
		\begin{tabular}{ccccccccc}
			\hline
			\noalign
			{\smallskip}
			\textrm {HJD'} & \textrm {$\phi_{o}$}  &\textrm{RV (km\,s$^{-1}$)} &\textrm {HJD'} & \textrm {$\phi_{o}$}  &\textrm{RV (km\,s$^{-1}$)}&\textrm {HJD'} & \textrm {$\phi_{o}$}  &\textrm{RV (km\,s$^{-1}$)} \\
			\hline
			\hline
			\noalign{\smallskip}
			\textrm{819.69328714} & \textrm{0.1477} & \textrm{-31.7} &\textrm{742.68658565} & \textrm{0.6559} & \textrm{ 52.3} &\textrm{938.50247685} & \textrm{0.8778} & \textrm{ 30.9} \\
			\textrm{565.32734954} & \textrm{0.1883} & \textrm{-23.4} &\textrm{937.46894670} & \textrm{0.7236} & \textrm{ 39.1} &	\textrm{938.51725694} & \textrm{0.8800} & \textrm{ 13.6} \\
			\textrm{565.33864583} & \textrm{0.1900} & \textrm{-32.5} &	\textrm{937.49209491} & \textrm{0.7270} & \textrm{ 10.8} &\textrm{938.53203704} & \textrm{0.8822} & \textrm{ 10.7} \\
			\textrm{820.68160635} & \textrm{0.2952} & \textrm{-25.0} &	\textrm{937.50343750} & \textrm{0.7287} & \textrm{ 40.6} &\textrm{563.32796296} & \textrm{0.8900} & \textrm{  2.7} \\
			\textrm{968.45359954} & \textrm{0.3475} & \textrm{-21.6} &\textrm{562.35005787} & \textrm{0.7440} & \textrm{ 20.9} &	\textrm{563.34273148} & \textrm{0.8922} & \textrm{ 23.8} \\
			\textrm{968.46481481} & \textrm{0.3491} & \textrm{-18.1} &	\textrm{609.32633102} & \textrm{0.7544} & \textrm{ 35.9} &	\textrm{563.35752315} & \textrm{0.8944} & \textrm{ 13.5} \\
			\textrm{741.67127315} & \textrm{0.5044} & \textrm{  9.3} &\textrm{743.66797454} & \textrm{0.8024} & \textrm{ 12.6} &	\textrm{610.29773148} & \textrm{0.8993} & \textrm{  8.4} \\
			\textrm{741.68372685} & \textrm{0.5063} & \textrm{ 17.6} &\textrm{938.47243056} & \textrm{0.8733} & \textrm{-47.3} &	\textrm{610.30902778} & \textrm{0.9010} & \textrm{ 14.1} \\
			\textrm{822.69305410} & \textrm{0.5954} & \textrm{ 23.4} &	\textrm{938.48769676} & \textrm{0.8756} & \textrm{ 17.0} & & & \\
			\hline
		\end{tabular}
		
	\end{center}
	\label{Tab: Tab. 7}
\end{table*}


\subsection{Spectral disentangling}
\label{Subsec: Subsec. 4.2}

\begin{figure}
	\begin{center}
		\includegraphics[trim=0.3cm 0.2cm 0.2cm 0.2cm,clip,width=0.45\textwidth,angle=00]{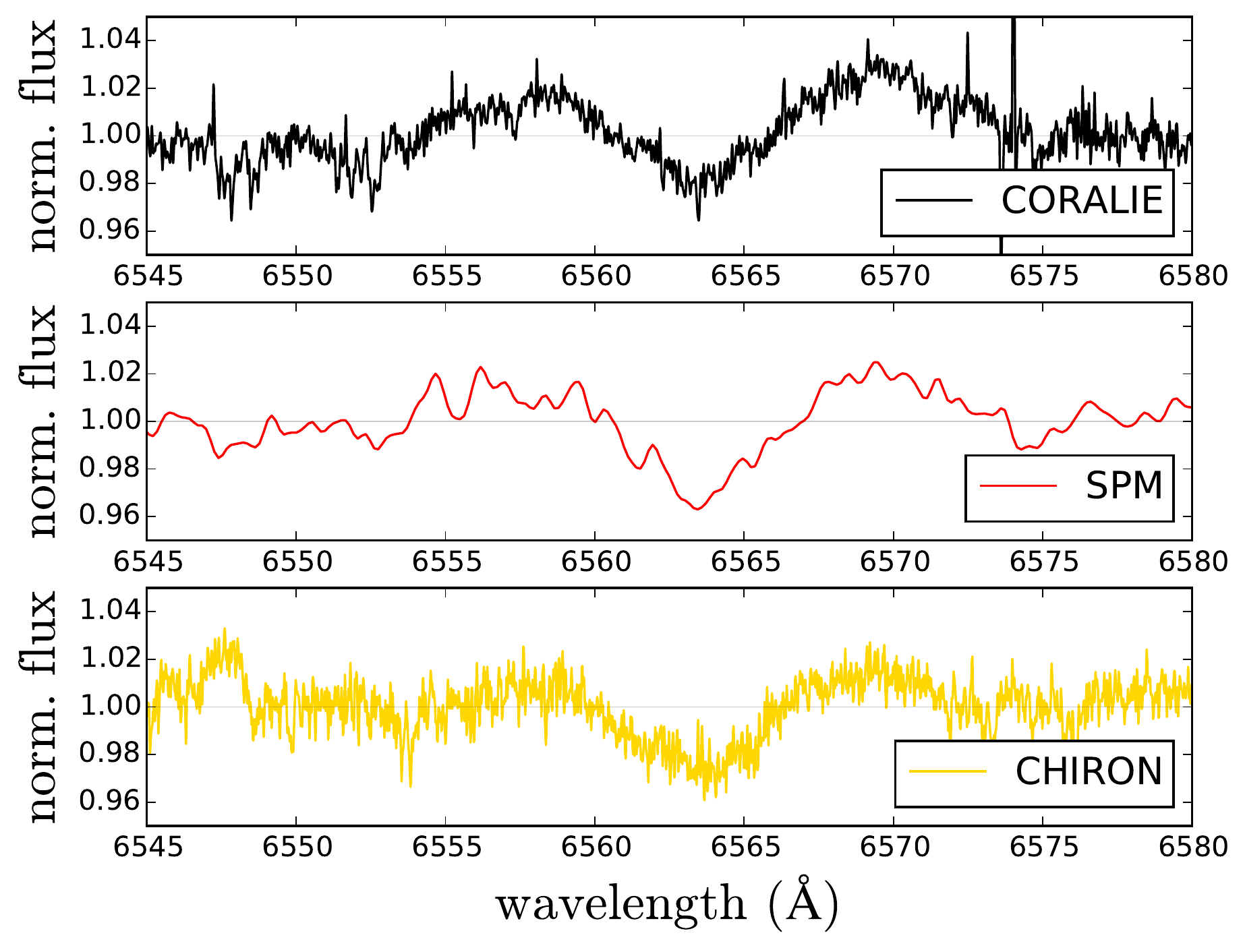}
	\end{center}
	\caption{Disentangled spectra of the gainer star around the H$\alpha$ line. Data of three different spectrographs are independently shown.}
	\label{fig:Fig. 4}
\end{figure}

Since we have independently distinguished the absorption lines of each component using the CORALIE spectra, we performed an iterative method of spectral disentangling proposed by \citet{2006A&A...448..283G}, which is quite effective in separating the absorption lines of the stellar components. This method uses
alternately the spectrum of one component to calculate the spectrum of the other one, hence eliminating gradually the spectral features of one stellar component until the convergence is assured and virtually disappears the flux contribution of its companion. In the process we used the theoretical radial velocities obtained from the sinusoidal fits in the previous section as  input parameters until the seventh iteration for both components, obtaining successfully clean average spectra for both stars.  The disentangling process for the gainer star reveals a double emission line in H$\alpha$, confirming the presence of circumstellar matter around the hot component. Furthermore, as the emission is double, an accretion disc is in principle inferred, in agreement with a semidetached
algol mass-transferring binary (Fig. \ref{fig:Fig. 4}).

\subsection{Determination of the donor physical parameters}
\label{Subsec: Subsec. 4.3}

The determination of physical parameters of the donor is of fundamental importance to constrain the size of the system. We have constructed a grid of synthetic theoretical spectra, modeling stellar atmospheres with the code \texttt{SPECTRUM} \footnote{\url{http://www.appstate.edu/~grayro/spectrum/spectrum.html}} \citep{1994AJ....107..742G,2001AJ....121.2159G} and using a grid of model atmospheres given by ATLAS9\footnote{\url{http://wwwuser.oats.inaf.it/castelli/grids.html}} \citep{2003IAUS..210P.A20C}. The grid was constructed in Local Thermodynamic Equilibrium (LTE) with different free parameters such as the effective temperature, which has two groups which vary $4000 \leq T_\mathrm{2} < 11000 ~\mathrm{K}$ with steps of 250 K and $11000 \leq T_\mathrm{2} \leq 20000 ~\mathrm{K}$ with steps of 1000\,K. The surface gravities vary from 0.0 to 5.0 dex with steps of 0.5 dex, the micro-turbulent velocity which is defined as the scale of turbulence in the stellar atmosphere in which the size of a turbulent cell is smaller than the mean free path, has unique options of 0.0 and 2.0 ~$\mathrm{km\,s}^{-1}$, $v\sin{i}$ varies from 0 to 150 ~$\mathrm{km\,s}^{-1}$ with step of 10 ~$\mathrm{km\,s}^{-1}$ and the macro-turbulence velocity from 0 to 10 ~$\mathrm{km\,s}^{-1}$ with step 1 ~$\mathrm{km\,s}^{-1}$. This parameter is defined as the scale of turbulence in the stellar atmosphere in which the size of the turbulence cell is greater than the mean free path of the photon. We also consider a veiling factor from 0.10 to 0.90 with step of 0.1 (dimensionless), this is defined as a constant of proportionality between the theoretical spectrum and the observed one, needed to account for the light contribution of the hot star veiling the absorption lines of the more evolved and cold star.  Finally, a fixed mixing length parameter $l/\mathrm{H}= 1.25$ was the default for the grid, this parameter corresponds when the blobs of convected fluid travel a distance {\textquotedblleft $l$\textquotedblright} from their position of equilibrium and then disrupt and disperse into the surroundings. 

\begin{figure}
	\begin{center}
		\includegraphics[trim=2.5cm 1.5cm 2.5cm 2.8cm,clip,width=0.45\textwidth,angle=0]{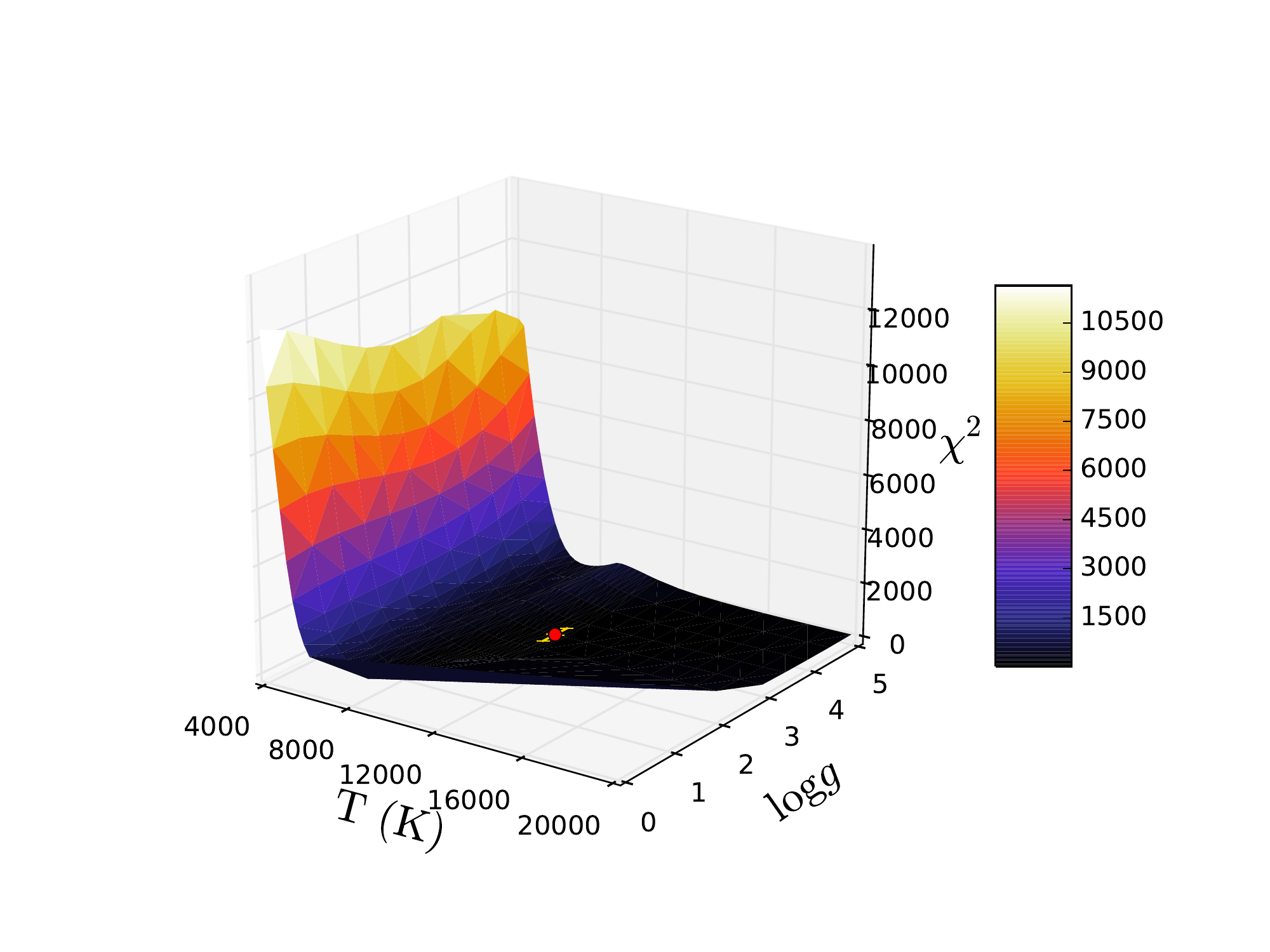}
	\end{center}
	\caption{Chi-square analysis to find the best theoretical normalized spectrum with respect to the average observed spectrum, computed with six free degrees obtained for the cold star, and with all parameters at their optimized values. The best model is obtained at $\chi2=14.197$ and is represented by a red dot at $T=10500 ~\mathrm{K}$ and a $\log{g}=3.0 ~\mathrm{dex}$.}
	\label{fig:Fig. 5}
\end{figure}

\begin{figure}
	\begin{center}
		\includegraphics[trim=0.0cm 0.0cm 0.3cm 0.2cm,clip,width=0.45\textwidth,angle=0]{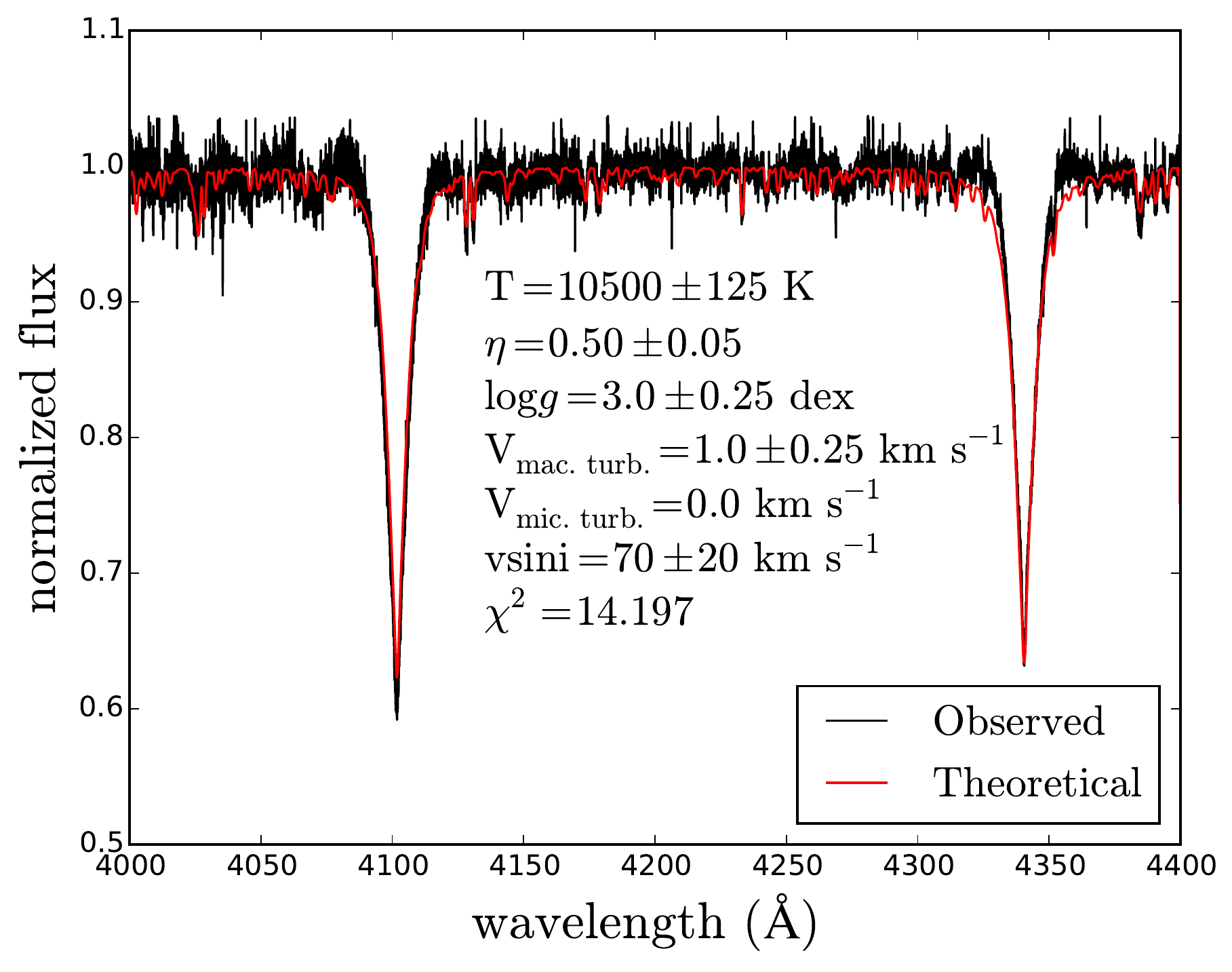}
	\end{center}
	\caption{Comparison between the best theoretical model obtained for the cold star (red line) and the disentangled average observed spectrum (black line).}
	\label{fig:Fig. 6}
\end{figure}

\begin{figure*}
	\begin{center}
		\includegraphics[trim=0.3cm 0.3cm 0.2cm 0.3cm,clip,width=1.0\textwidth,angle=0]{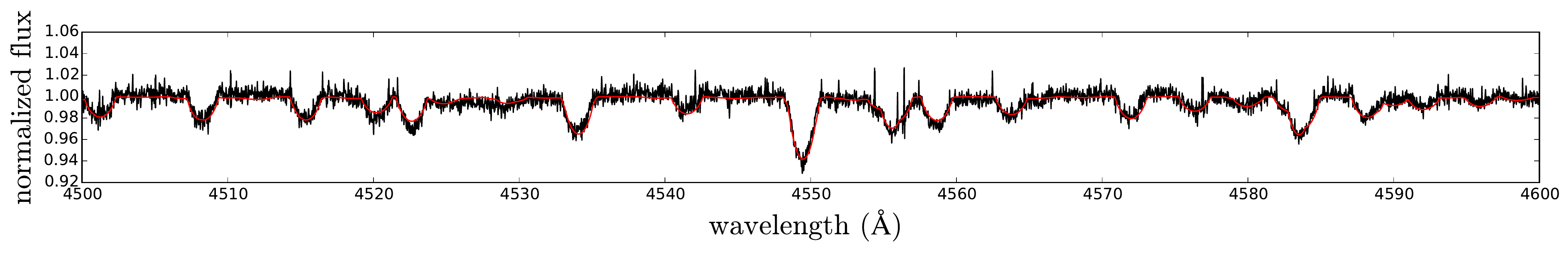}
	\end{center}
	\caption{Detailed comparison between the disentangled observed (black line) and theoretical (red line) secondary spectrum.}
	\label{fig:Fig. 7}
\end{figure*}

The implemented analysis was based on a chi-square optimization algorithm $\chi2=\sum_{i=\lambda_{0}}^{\lambda_\mathrm{f}}(O_\mathrm{i}-E_\mathrm{i})^{2}/E_\mathrm{i}$, this consists in the minimization of the deviation between the theoretical normalized spectrum ($E_\mathrm{i}$) and the observed average spectrum of the donor star already disentangled and corrected by a veiling factor on the analyzed spectral range ($O_\mathrm{i}$), and the lines were analyzed at the rest frame. 
The implemented method yielded the best result with a temperature of $T_{2}= 10500 \pm 125 ~\mathrm{K}$, $\log{\mathrm{g}}_{2}=3.0 \pm 0.5 ~\mathrm{dex}$, with fixed $v_\mathrm{mic}=0.0 ~\mathrm{km\,s^{-1}}$, $v_\mathrm{mac}= 1.0 ~\mathrm{km\,s^{-1}}$, $v\sin{i}= 70 \pm 20 ~\mathrm{km\,s^{-1}}$, veiling factor $\eta = 0.5 \pm 0.05$ and a $l/\mathrm{H}=1.25$ (Fig. \ref{fig:Fig. 5}). The comparison between the observed and the theoretical spectrum shows a good match in the 4000-4400 \AA\ region as well as in the broader region of
4500-4600 \AA\ (Fig.\,\ref{fig:Fig. 6} and Fig.\,\ref{fig:Fig. 7}). 

An inspection of the behaviour of $\chi2$ versus $v\sin{i}$ shows that the projected rotational velocity is loosely constrained. Actually, $\chi2$
changes very little in the range 50-150  $\mathrm{km\,s}^{-1}$.
If we assume synchronous rotation for the donor, because of the close binary nature \citep{1975A&A....41..329Z,1977A&A....57..383Z}, and considering that the expected rotational velocity can be calculated from equation 3.9 of  \citet{2006epbm.book.....E},

\begin{equation}
	\frac{v_\mathrm{rot}\sin{i}}{K_2} = \frac{R_\mathrm{L}}{a} \approx (1+q)\frac{0.49q^{2/3}}{ 0.6q^{2/3}+ \ln(1+q^{1/3})},
	\label{eq: eq. 8}
\end{equation}

\noindent
where $R_\mathrm{L}$ is the volume Roche Lobe radius of the secondary, we get $v\sin{i}$ = 47 $\mathrm{km\,s}^{-1}$, using $q$ = 0.206 and $K_{2}= 154 ~\mathrm{km\,s^{-1}}$. This is just in the lower limit of our determined value for the donor projected rotational velocity.
Therefore, our analysis is consistent with a rotationally synchronized donor star. We notice that the approximation in Eq.\,\ref{eq: eq. 8} is accurate to 1\% for all $q$ values \citep{2006epbm.book.....E}. 

We calculated $R_\mathrm{L}/a= 0.455 \pm 0.130$ using Eq.\,\ref{eq: eq. 8}. Assuming $R_2$ = $R_\mathrm{L}$ we have quantified the mean density $\bar{\rho}$ (g\,cm$^{-3}$) of the donor star. Using $\rho=3M_{2}/4\pi R^{3}_{\mathrm{2}}$, $M_{\mathrm{2}}=Mq/(1+q)$ and the Kepler's third law we get:

\begin{equation}
	\bar{\rho}_{2}= \frac{3q}{(1+q)}\frac{1}{(R_{\mathrm{2}}/a)^{3}}\frac{\pi}{G P^{2}},
	\label{eq: eq. 9}
\end{equation}

\noindent
and we obtain a highly affected mean density of $\bar{\rho}_{2}=0.0008 \pm 0.0004 ~\mathrm{g\,cm^{-3}}$, which is characteristic of evolved stars, possibly caused by the loss mass during the mass transfer process.

\subsection{On the gainer star}

\begin{figure}
	\begin{center}
		\includegraphics[trim=0.2cm 0.2cm 0.2cm 0.2cm,clip,width=0.45\textwidth,angle=0]{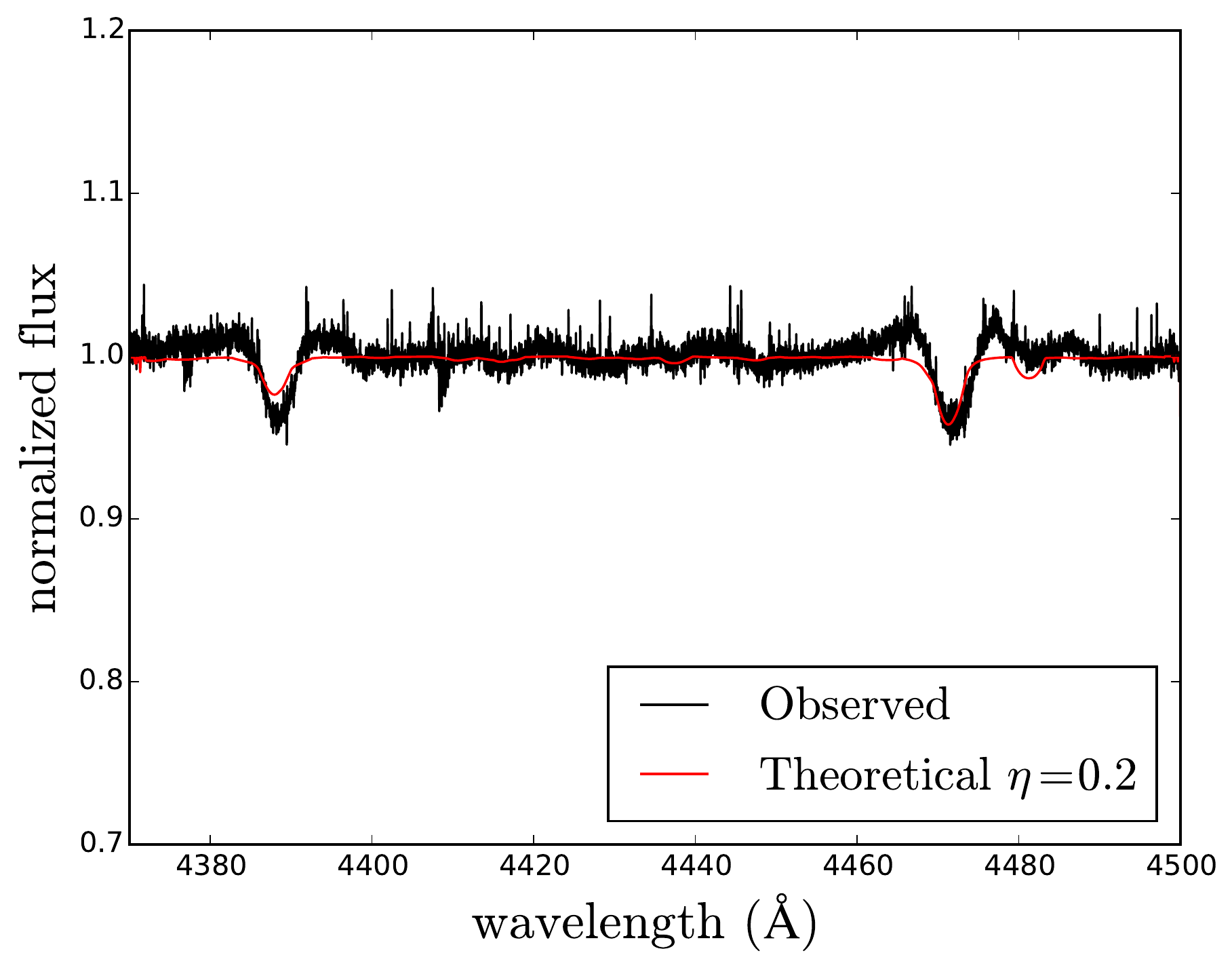}
	\end{center}
	\caption{Disentangled CORALIE spectrum for the gainer star, showing broad helium lines surrounded by emission shoulders and the best model.}
	\label{fig:Fig. 8}
\end{figure}

A similar procedure as described in the previous section was performed for the gainer star. The spectrum of the gainer is characterized by weak H$\alpha$ emission and weak H$\beta$, H$\gamma$ and H$\delta$ absorptions, along with broad He\,I lines, sometimes surrounded by emission flanks. This suggest that the spectrum is a blend of the contribution of the gainer and the accretion disk.  We choose a spectral range with only helium lines to search for the best spectral model, and run the $v\sin{i}$ parameter between 10 and $400 ~\mathrm{km\,s^{-1}}$. We find a best model with $T_{1}= 16000 ~\mathrm{K}$, $\log{g} \geq 3.0 ~\mathrm{dex}$, $v_\mathrm{mic}= 0.0 ~\mathrm{km\,s^{-1}}$, $v\sin{i}= 130 ~\mathrm{km\,s^{-1}}$, $v_\mathrm{mac}= 1 ~\mathrm{km\,s^{-1}}$, veiling factor $\eta= 0.2$ and $\chi2= 2.215$ (Fig. \ref{fig:Fig. 8}). This suggests that the gainer is a B-type star with spectral type around B\,4. We notice that the low surface gravity found could be due to the influence of line emission of circumstellar material affecting the width of the line profiles.

\subsection{The H$\alpha$ and H$\beta$ emission line profiles}
\label{Subsec: Subsec. 4.5}

\begin{figure}
\begin{center}
\includegraphics[trim=0.2cm 0.2cm 0.2cm 0.2cm,clip,width=0.45\textwidth,angle=0]{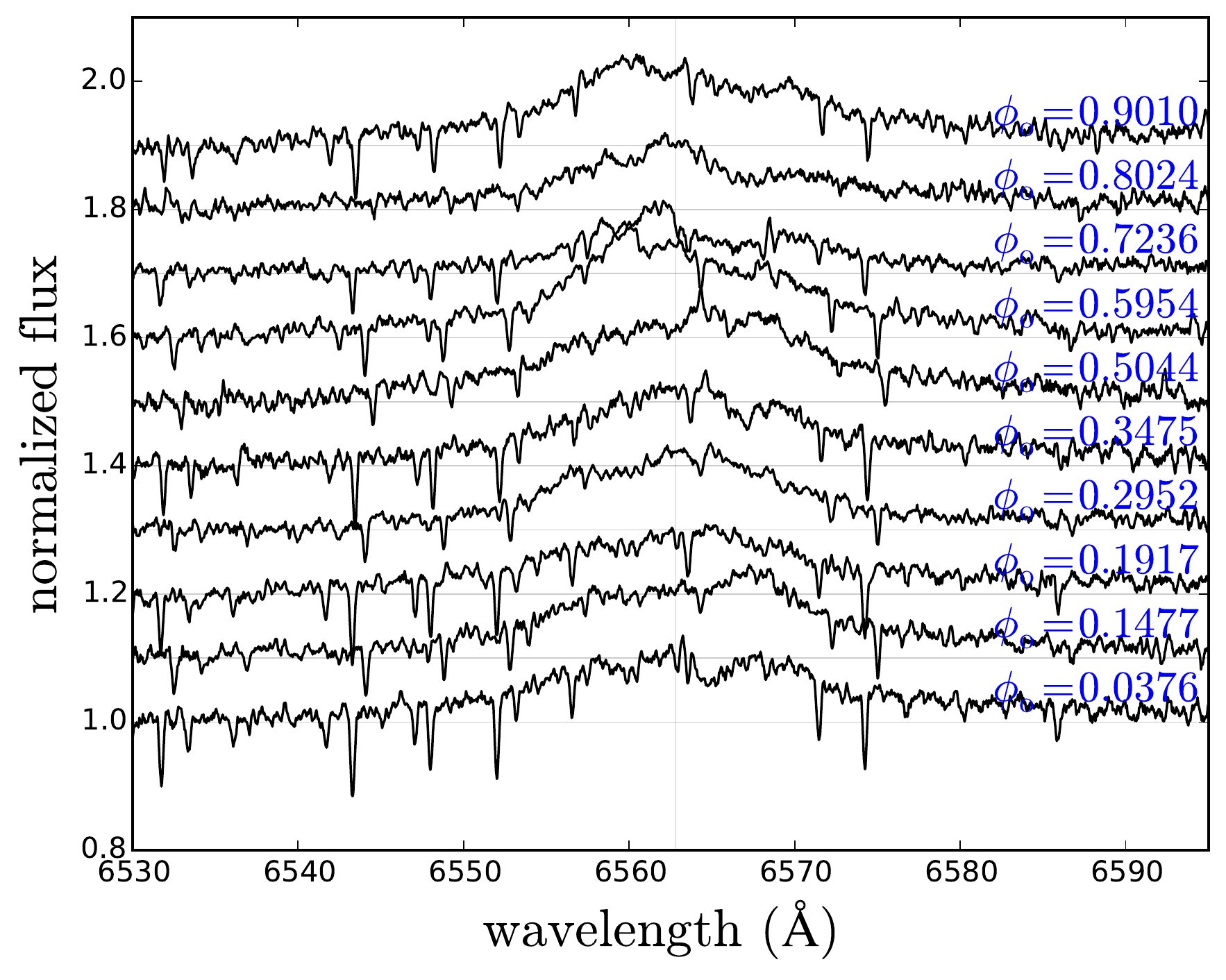}
\includegraphics[trim=0.2cm 0.2cm 0.2cm 0.2cm,clip,width=0.45\textwidth,angle=0]{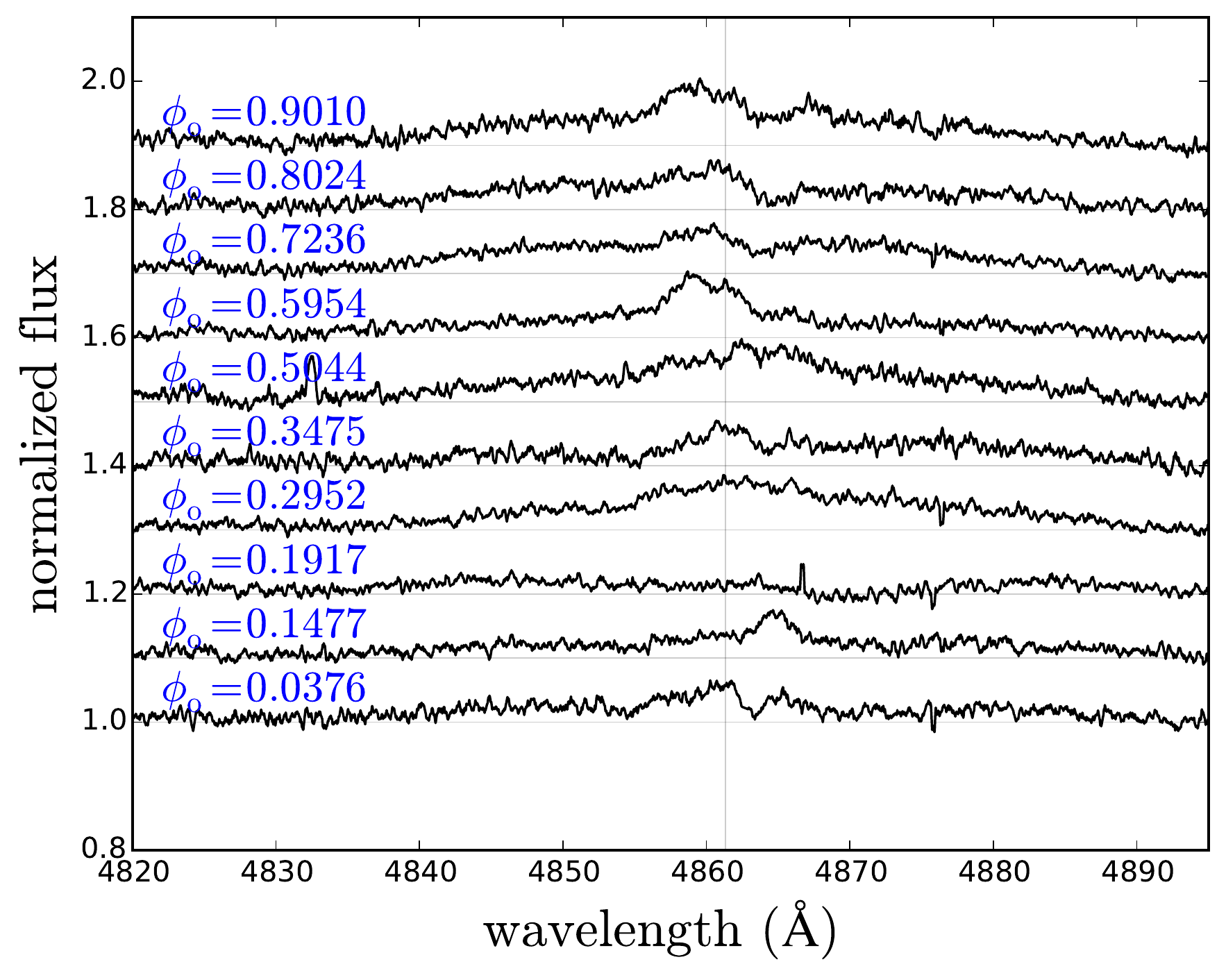}
\end{center}
\caption{Residual H$\alpha$ (up) and H$\beta$ (down) emission in CORALIE spectra, smoothed by a factor 50, during a complete orbital cycle. The orbital phases are displayed.}
\label{fig:Fig. 9}
\end{figure}

\begin{figure}
	\begin{center}
		\includegraphics[trim=0.2cm 0.2cm 0.2cm 0.2cm,clip,width=0.45\textwidth,angle=0]{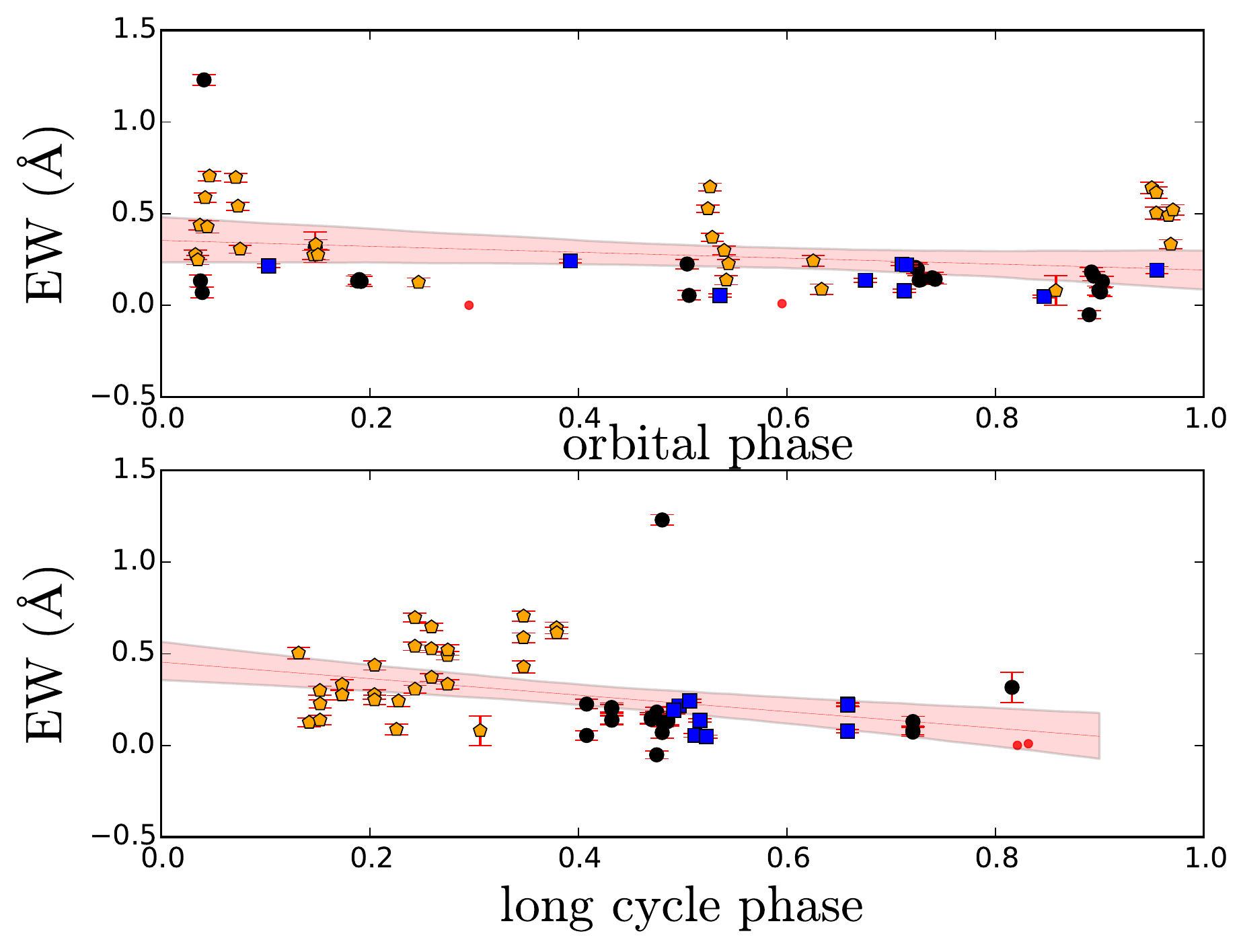}
	\end{center}
	\caption{Equivalent width of the H$\alpha$  profile of the gainer star for several orbital (Top) and long cycle phases (Bottom) without the donor contribution, using CORALIE (black dots), Echelle (blue dots) and CHIRON (orange dots) spectrographs. The red continuum line is a fitted 1st order polynomial to the EWs with a confidence interval region of $95\%$.}
	\label{fig:Fig. 10}
\end{figure}

We have performed a complementary analysis to the H${\alpha}$ and H$\beta$ profiles removing  the donor and gainer theoretical contributions. We observe H${\alpha}$ and H$\beta$ emission at all phases, being in H$\alpha$ broad and usually double (Fig. \ref{fig:Fig. 9}). This is consistent with an origin in an accretion disk. 
The profile shapes show variable asymmetry. 
The H$\alpha$ equivalent width - measured before gainer subtraction - changes through the orbital cycle  but most notably during the long cycle, being lower (eventually due to more emission) near the long-cycle maximum (Fig. \ref{fig:Fig. 10}). 

\begin{figure}
	\begin{center}
		\includegraphics[trim=0.2cm 0.2cm 0.2cm 0.2cm,clip,width=0.45\textwidth,angle=0]{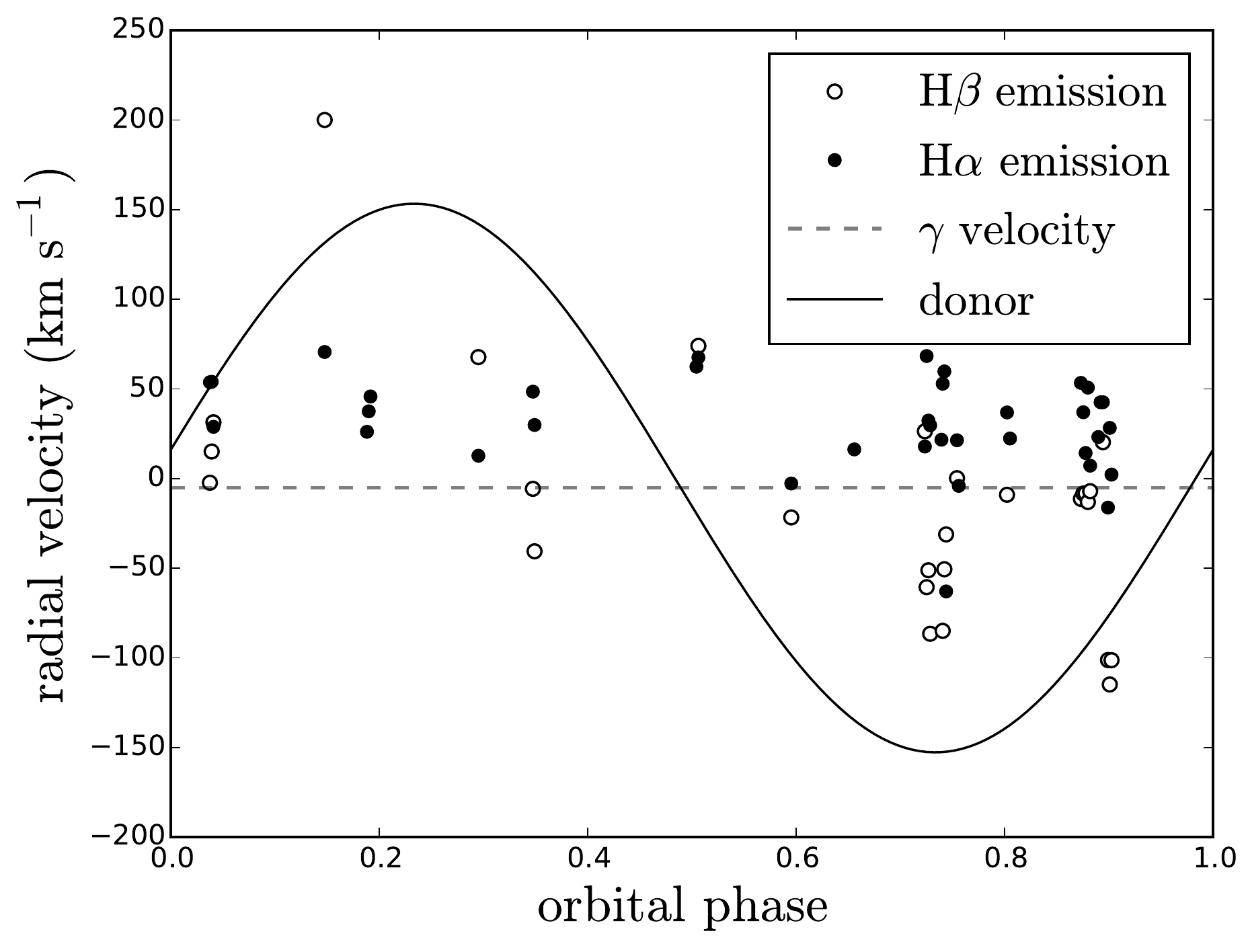}
	\end{center}
	\caption{Radial velocities of the residual emission along with the theoretical fit to the RVs of the donor.}
	\label{fig:Fig. 11}
\end{figure}

We measured the RV of the H$\alpha$ residual emissions using fits with Gaussian functions, but the H$\beta$ emission was quite irregular and in this case we measured the RVs using a fit only considering the upper portion of the profiles. The velocities show large scatter; the averages are $\mathrm{\overline{RV}}_{\alpha}= 30.6 \pm 26.7 ~\mathrm{km\,s^{-1}}$ and $\mathrm{\overline{RV}}_{\beta}= -42.3 \pm 85.2 ~\mathrm{km\,s^{-1}}$. While the H$\alpha$ velocities are consistent with no orbital variability, the H$\beta$ velocities seems to roughly follow the motion of the donor but with large scatter (Fig. \ref{fig:Fig. 11}).

\begin{figure*}
\begin{center}
\includegraphics[trim=1.0cm 0.3cm 2.0cm 1.0cm,clip,width=0.33\textwidth,angle=0]{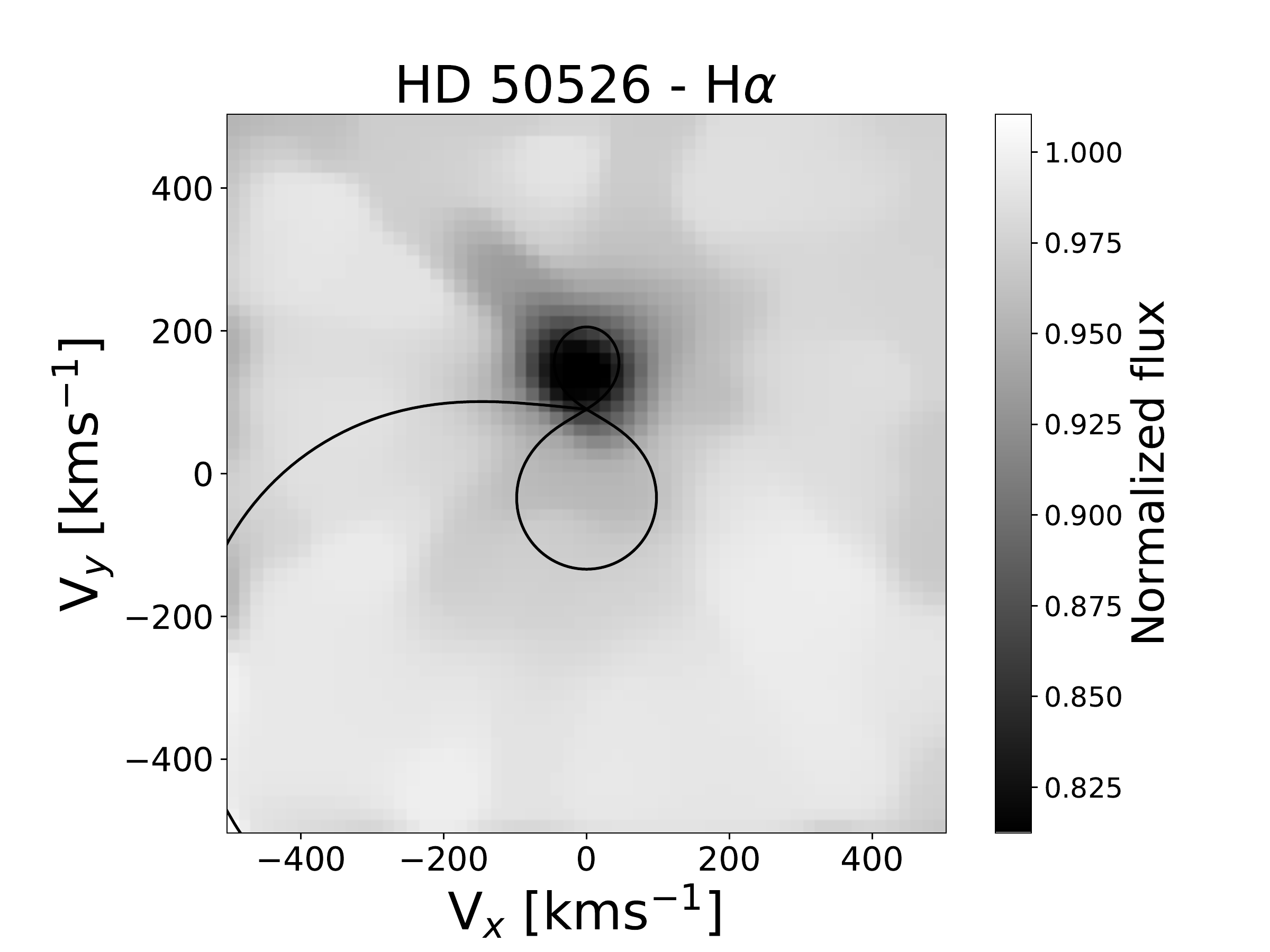}
\includegraphics[trim=1.0cm 0.3cm 2.0cm 1.0cm,clip,width=0.33\textwidth,angle=0]{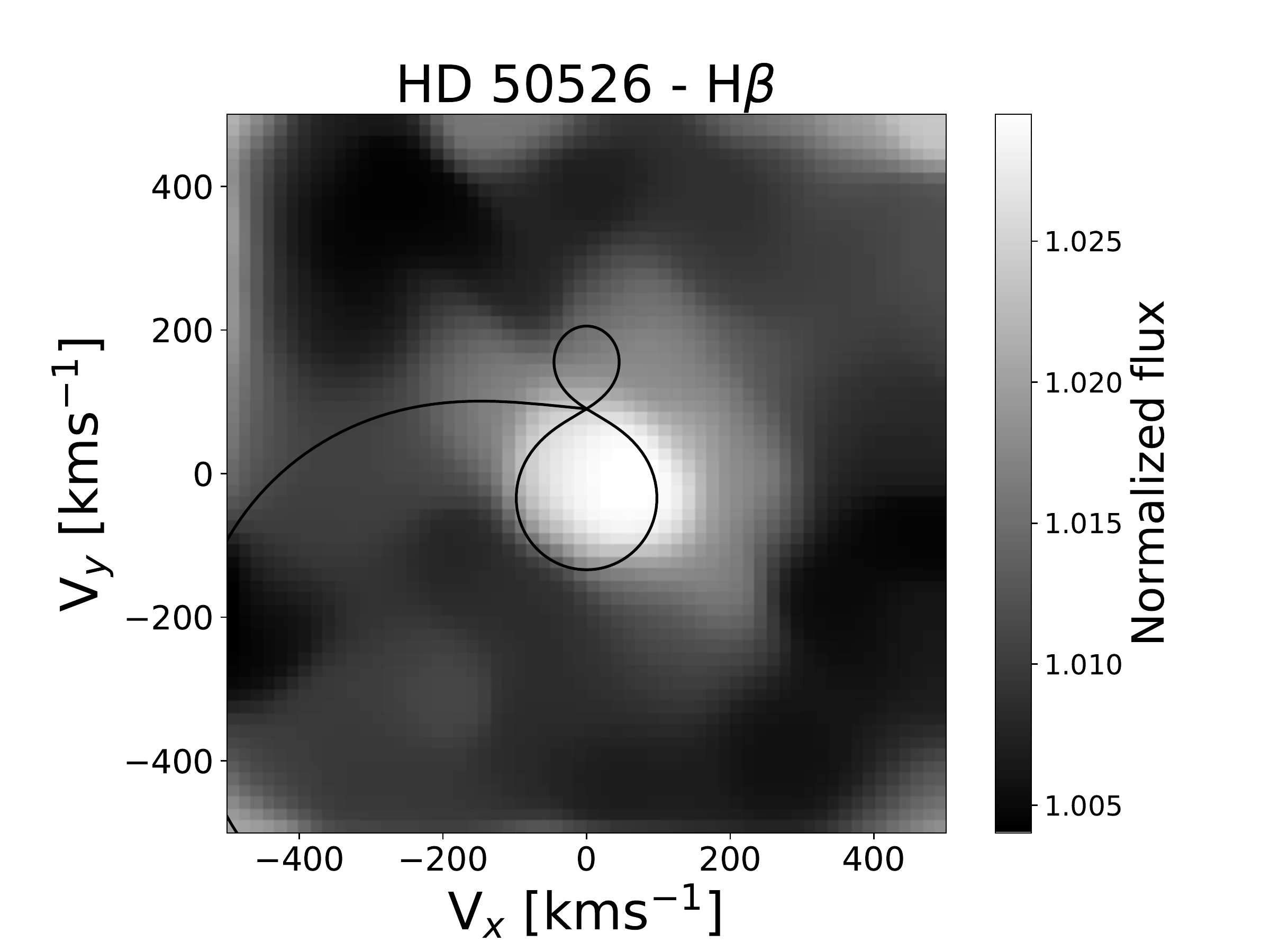}
\includegraphics[trim=1.0cm 0.3cm 2.0cm 1.0cm,clip,width=0.33\textwidth,angle=0]{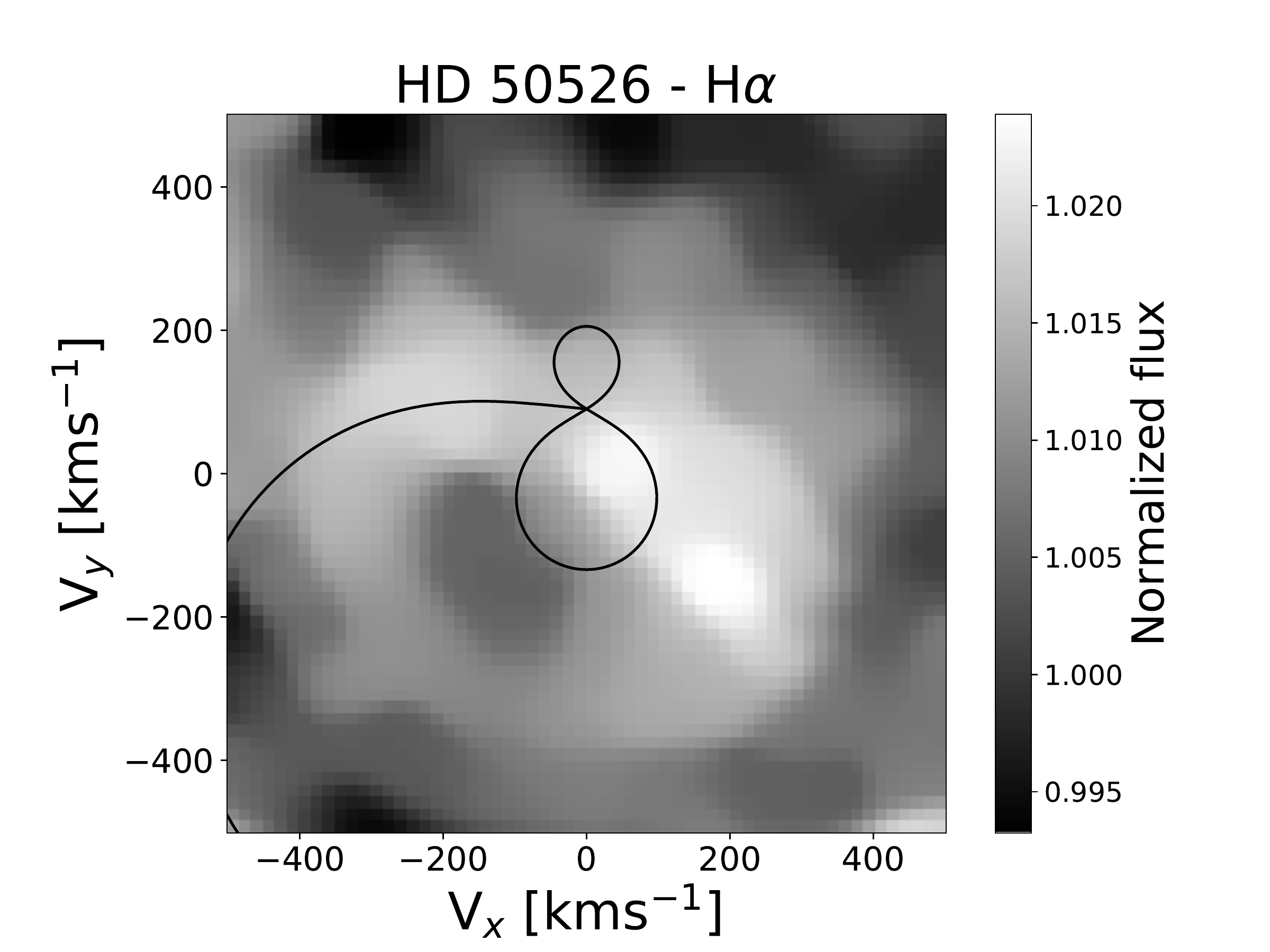}
\end{center}
\caption{Doppler maps of disentangled spectra: donor H$\alpha$ absorption (left) and H$\alpha$ and H$\beta$ emission (donor and gainer subtracted, center and right). Flux 1 indicates the continuum level. The Roche lobes of the donor and gainer are plotted projected in the orbital plane, along with the theoretical path followed by the gas stream. }
	\label{fig:Fig. 12}
\end{figure*}


\subsection{Doppler tomography}

Doppler tomography is an image reconstruction technique introduced by \citet{1988MNRAS.235..269M} that has been widely used in the study of cataclysmic variables and more recently, in the study of algol systems \citep{2004AN....325..229R}. The technique  consists of the reconstruction of the system emissivity in the velocity space from spectra taken along the orbital cycle. 

We constructed Doppler maps for the H$\alpha$ and H$\beta$ lines using 25 CHIRON spectra, 37 CORALIE spectra and 15 SPM spectra.
These were selected as those spectra with highest S/N sampling well the orbital cycle and we used the disentangled spectra, alternatively removing the contribution of the donor and gainer star. The Doppler maps  were constructed by means of the Total Variation Minimization (DTTVM) code that also deals with absorption lines; this code was presented by \citet{2015PASJ...67...22U}.

We find that the H$\alpha$ donor spectra reveal an absorption line centered in the cold star, as expected (Fig.\,\ref{fig:Fig. 12}, left panel). In addition, the residual emission is observed inside the Roche lobe of the gainer, much clearly in the H$\beta$ line without further structure, while the H$\alpha$ line shows additional high-velocity emission structures in the lower right quadrant and a diffuse and thick donuts-shaped emission structure centered at the low-left side of the gainer (Fig.\,\ref{fig:Fig. 12}, central and right panels). An additional mild emission is observed in the upper left quadrant around the track of the gas stream in the H$\alpha$ map. These emissions are reminiscent of that expected from an optically thin accretion disk observed at mid latitudes. However, the presence of structured regions of higher emissivity and the localized H$\beta$ emission inside the Roche lobe of the gainer probably reflects the complex geometrical and physical structure of the circumstellar material surrounding the gainer. This material might even not to be restricted to the orbital plane, breaking the assumptions of the Doppler tomography reconstruction. Eventual mass outflows would also influence the general appearance of the Doppler maps.

\section{Light-curve model and system parameters}
\label{Sec: Sec. 5}

In this section we determine the stellar and system parameters that best fit the orbital light curve.

\subsection{The fitting procedure}
\label{Subsec: Subsec. 2.1}

We fit the orbital light curve with a code described by  \citet{1992Ap&SS.197.17D}. The algorithm applies the inverse-problem solving method based on the simplex algorithm, considering a binary system with an optically thick disc surrounding the more massive star. The system is assumed semidetached, with the less massive star filling its Roche lobe. We used the Nelder-Mead simplex algorithm \citep{1992nrfa.book.....P}
with optimizations described by \citet{DT91}.  While the direct problem comprises the calculation of the light curve from model parameters
given {\it a priori}, the inverse problem is the process of finding the set of parameters that will optimally fit the synthetic light curve to the observations.  

The theoretical model considers a hot spot located on the edge of the disc, in the place where the gas stream from the donor falls encountering the disc. This active region is described by the ratio of the hot spot temperature and the unperturbed local disc temperature and the angular dimension and longitude of the spot. An additional, bright spot, with similar parameters, is also included in the outer disc edge, 
following results of hydro-dynamical simulations of gas interchanged among close binary systems \citep[e.g.][]{1994A&A...288..807H, 2017ARep...61..639K}. The disk is assumed in physical contact with the gainer and is characterized by its radius Rd, outer and inner edge thicknesses and its temperature:

\begin{equation}
	T(r) = T_\mathrm{d} \left(\frac{T_\mathrm{d}}{r}\right)^{a_\mathrm{T}}
\end{equation}

\noindent
where $T_\mathrm{d}$ is the disk temperature at its outer edge ($r=R_\mathrm{d}$) and $a_\mathrm{T}$ is the temperature exponent ($a_\mathrm{T} \leq 0.75$). The value of exponent $a_\mathrm{T}$ shows how close is the radial temperature profile to the steady-state configuration ($a_\mathrm{T}=0.75$). The radial dependencies of the accretion disk temperature shows that the surface of the disk is hotter in the inner regions, and that disk gets cold as one moves away from the center. The model and code have been widely used during our recent research of intermediate-mass interacting binaries \citep{2013MNRAS.432..799M, 2015MNRAS.448.1137M, 2018MNRAS.476.3039R}. Based on our results of previous sections, we fixed the mass ratio to $q = 0.206$ and the temperature of the cold and less massive star to $T_{2}=10500 ~\mathrm{K}$.

\begin{table}
	\caption{Results of the analysis of {DPV HD\,50526} V-filter light-curve obtained by solving the inverse problem for the Roche model with an accretion disk around the more-massive (hotter) gainer in the critical non-synchronous rotation regime.}
	\label{Tab: Tab. 8}
	\normalsize
	\resizebox{0.4\textwidth}{5.5cm}{$
	\begin{array}{llll}
	\hline
	\noalign{\smallskip}
	{\rm Quantity} & \  \  & {\rm Quantity} & \\
	\noalign{\smallskip}
	\hline
	\noalign{\smallskip}
	n                               & 265   \  \     & \cal M_{\rm_1} {[\cal M_{\odot}]} & 5.48 \pm 0.02\\
	{\rm \Sigma(O-C)^2}                & 0.0565\  \     & \cal M_{\rm_2} {[\cal M_{\odot}]} & 1.13 \pm 0.02\\
	{\rm \sigma_{rms}}                 & 0.0146\  \     & \cal R_{\rm_1} {\rm [R_{\odot}]}  & 3.57 \pm 0.03\\
	i {\rm [^{\circ}]}              & 61.28 \pm 0.2  & \cal R_{\rm_2} {\rm [R_{\odot}]}  & 7.09 \pm 0.01\\
	{\rm F_d}                          & 0.998 \pm 0.2  & {\rm log} \ g_{\rm_1}             & 4.07 \pm 0.02\\
	{\rm T_d} [{\rm K}]                & 9401  \pm 200  & {\rm log} \ g_{\rm_2}             & 2.79 \pm 0.02\\
	{\rm d_e} [a_{\rm orb}]            & 0.152 \pm 0.01 & M^{\rm h}_{\rm bol}               &-2.40 \pm 0.26\\
	{\rm d_c} [a_{\rm orb}]            & 0.102 \pm 0.02 & M^{\rm 2}_{\rm bol}               &-2.06 \pm 0.02\\
	{\rm a_T}                          & 0.41  \pm 0.05 & a_{\rm orb}  {\rm [R_{\odot}]}    & 28.04\pm 0.02\\
	{\rm f_1}                          & 14.71 \pm 0.5  & \cal{R}_{\rm d} {\rm [R_{\odot}]} & 14.74\pm 0.02\\
	{\rm F_1}                          & 1.000  \  \    & \rm{d_e}  {\rm [R_{\odot}]}       & 4.26 \pm 0.02\\
	{\rm T_1} [{\rm K}]                & 16000  \  \    & \rm{d_c}  {\rm [R_{\odot}]}       & 2.87 \pm 0.03\\
	{\rm T_2} [{\rm K}]                & 10500  \  \    &                                                  \\
	{\rm A_{hs}=T_{hs}/T_d}            & 1.36  \pm 0.03 &                                                  \\
	{\rm \theta_{hs}}{\rm [^{\circ}]}  & 19.8  \pm 2.0  &                                                  \\
	{\rm \lambda_{hs}}{\rm [^{\circ}]} & 343.1 \pm 3.0  &                                                  \\
	{\rm \theta_{rad}}{\rm [^{\circ}]} & -16.2 \pm 2.0  &                                                  \\
	{\rm A_{bs}=T_{bs}/T_d}            & 1.18  \pm 0.03 &                                                  \\
	{\rm \theta_{bs}}{\rm [^{\circ}]}  & 46.6  \pm 4.0  &                                                  \\
	{\rm \lambda_{bs}}{\rm [^{\circ}]} & 129.5 \pm 5.0  &                                                  \\
	{\Omega_{\rm 1}}                   & 9.796 \pm 0.09 &                                                  \\
	{\Omega_{\rm 2}}                   & 2.248 \  \     &                                                  \\
	\noalign{\smallskip}                    
	\hline
	
	\end{array}
	$}
	
	\small FIXED PARAMETERS: $q={\cal M}_{\rm 2}/{\cal M}_{\rm 1}=0.206$ - mass ratio of the components, ${\rm F_1}=R_1/R_{zc}=1$ - filling factor for the critical Roche lobe of the hotter, more-massive gainer (ratio of the stellar polar radius to the critical Roche lobe radius along z-axis for a star in synchronous rotation regime), ${\rm T_1= 16000 K}$, ${\rm T_2= 10500 K}$  - temperature of the more massive (hotter) gainer and the less-massive (coldest) donor, ${\rm F_2}=1.0$ - filling factor for the critical Roche lobe of the donor, $f{\rm _{1,2}}= 1.00$ - non-synchronous rotation coefficients of the system components, ${\rm\beta_1= 0.25}$, ${\rm \beta_2= 0.25}$ - gravity-darkening coefficients of the components, ${\rm A_1= 1.0}$, ${\rm A_2= 1.0}$  - albedo coefficients of the components.
	
	\small \noindent Note: $n$ - number of observations, ${\rm \Sigma (O-C)^2}$ - final sum of squares of residuals between observed (LCO) and synthetic (LCC) light-curves, ${\rm \sigma_{rms}}$ - root-mean-square of the residuals, $i$ - orbit inclination (in arc degrees), ${\rm F_d=R_d/R_{yc}}$ - disk dimension factor (the ratio of the disk radius to the critical Roche lobe radius along y-axis), ${\rm T_d}$ - disk-edge temperature, $\rm{d_e}$, $\rm{d_c}$,  - disk thicknesses (at the edge and at the center of the disk, respectively) in the units of the distance between the components, $a_{\rm T}$ - alpha disk temperature distribution coefficient, $f{\rm _h}$ - non-synchronous rotation coefficient of the more massive gainer (in the synchronous rotation regime), ${\rm A_{hs}=T_{hs}/T_d}$ - hot spot temperature coefficient, ${\rm \theta_{hs}}$ and ${\rm \lambda_{hs}}$ - spot angular dimension and longitude (in arc degrees), ${\rm\theta_{rad}}$ - angle between the line perpendicular to the local disk edge surface and the direction of the hot-spot maximum radiation, ${\rm A_{bs}=T_{bs}/T_d}$ - 
	bright spot temperature coefficient, ${\rm \theta_{bs}}$ and ${\rm \lambda_{bs}}$ - bright spot angular dimension and longitude (in arc degrees), ${\Omega_{\rm 1,2}}$ - dimensionless surface potentials of the hotter gainer and coldest donor, $\cal M_{\rm_{1,2}} {[\cal M_{\odot}]}$, $\cal R_{\rm_{1,2}} {\rm [R_{\odot}]}$ - stellar masses and mean radii of stars in solar units, ${\rm log} \ g_{\rm_{1,2}}$ - logarithm (base 10) of the system components effective gravity, $M^{\rm {1,2}}_{\rm bol}$ - absolute stellar bolometric magnitudes, $a_{\rm orb}$ ${\rm [R_{\odot}]}$, $\cal{R}_{\rm d} {\rm [R_{\odot}]}$, $\rm{d_e} {\rm [R_{\odot}]}$,
	$\rm{d_c} {\rm [R_{\odot}]}$ - orbital semi-major axis, disk radius and disk thicknesses at its edge and center, respectively,
	given in solar units. 
\end{table}

\normalsize

\begin{figure}
	\begin{center}
		\includegraphics[trim=0.0cm 0.0cm 0.0cm 0.0cm,clip,width=0.37\textwidth,angle=0]{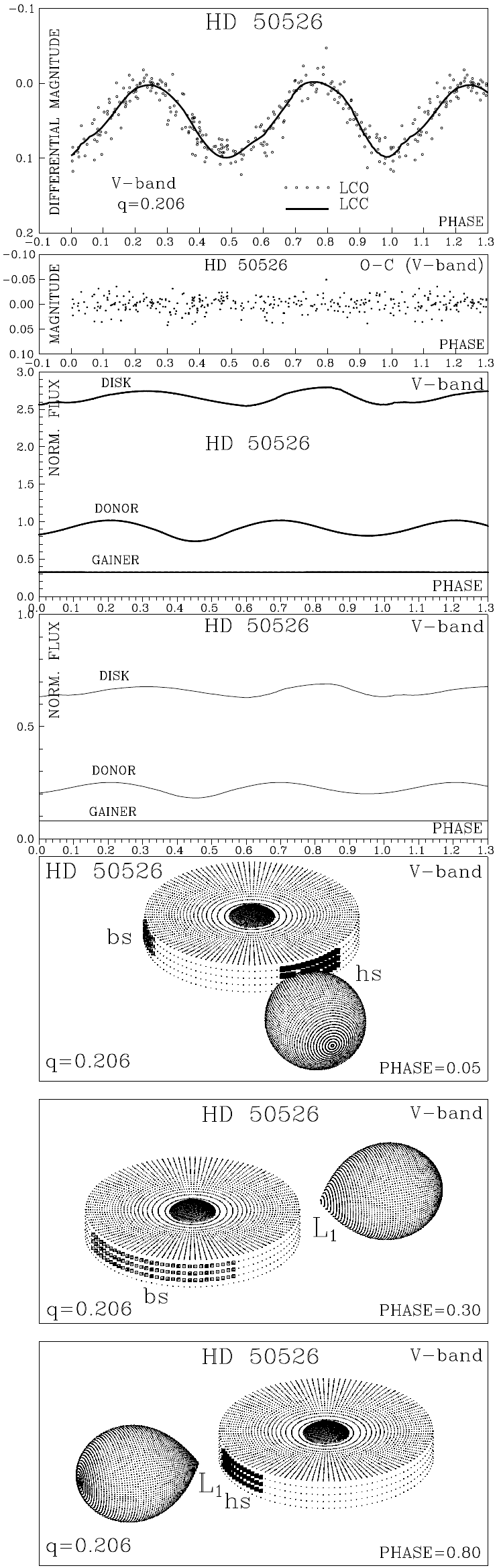}
	\end{center}
	\caption{Observed (LCO) and synthetic (LCC) light-curves of HD\,50526 obtained by analyzing photometric observations; final O-C residuals between the observed and synthetic light curves; fluxes of donor, gainer and of the accretion disc, normalized to the donor flux at phase 0.25; the views of the model at orbital phases 0.05, 0.30, and 0.80, obtained with parameters estimated by the light curve analysis.}
	\label{fig:Fig. 13}
\end{figure}


\subsection{The best light-curve model}
\label{Sub. Sec: Subsec. 5.2}

The obtained results for the system are summarized in Fig. \ref{fig:Fig. 13} and presented in Table \ref{Tab: Tab. 1}. The best model shows an orbital inclination $61.\!\!^{\circ}28 \pm 0.\!\!^{\circ}2$ degree. The stars have masses $5.48 \pm 0.02$ and $1.13 \pm 0.02  ~\mathrm{M_{\odot}}$, radii $3.57 \pm 0.03 ~\mathrm{R_{\odot}}$ and $7.09 \pm 0.01 ~\mathrm{R_{\odot}}$ and surface gravities $\log{g}=4.07 \pm 0.02 \mathrm{dex}$ and $\log{g}=2.79 \pm 0.02~\mathrm{dex}$. Considering their temperatures and surface gravities they correspond to a pair B5\,V + A0\,III. The B type star fits relatively well the mass-radius expectations for main sequence stars but the A type star turns to be slightly bigger than expected, which causes that its surface gravity and mean density appears affected and look more evolved than it appears to be.

The model also contains an accretion disc, concave and optically thick around the hot star with a radius of $R_\mathrm{d}= 14.74 \pm 0.02 ~\mathrm{R_{\odot}}$, which is 4.1 times larger than the star in its center ($R_\mathrm{h}=3.57\pm 0.031 ~\mathrm{R_{\odot}}$), with a central thickness $d_\mathrm{c}= 2.87 \pm 0.032 ~\mathrm{R_{\odot}}$ and an edge thickness $d_\mathrm{e}=4.26 \pm 0.02  ~\mathrm{R_{\odot}}$. The gainer star is partly hidden by the disc whose temperature in the external border is  $T_\mathrm{d}= 9401 \pm 200 ~\mathrm{K}$. The disk presents a hot spot on the its outer edge  with a temperature of $T_\mathrm{hs}= 12785 \pm 392 ~\mathrm{K}$ and a bright spot temperature of $T_\mathrm{bs}= 11093 \pm 368 ~\mathrm{K}$, placed almost in the opposite location of the disk. 

The disk dominates the light of the system at the $V$-band, contributing 67\% to the total flux at orbital phase 0.25, whereas the donor contributes 25\% and the gainer the remaining 8\%. Let's remember that the disk luminosity is not accretion-driven in double periodic variables, being the disk mainly heated by high-energy radiative flux from the hotter star  \citep{2016MNRAS.461.1674M}.

\section{Discussion}
\label{Sec: Sec. 6}

HD\,50526 turns to be a typical DPV, consisting of a B dwarf + A giant semidetached pair, and with evidence of circumstellar matter around the hotter star. While the continuum light curve is well represented including an optically thick accretion disk around the gainer, the line emission Doppler maps reflect the complex dynamics of gaseous mass fluxes.   

Spectroscopic studies of non-eclipsing double periodic variables are scarce. If the region causing the long photometric cycle is located somewhere in the orbital plane, it should be differently visible in eclipsing and non-eclipsing systems. In these later, the projected disk surface is larger and the observer should dispose of a better view of the disk features. It makes sense then to compare HD\,50526 with HD\,170582, two  DPVs seen under intermediate latitudes. Data for these two systems are listed in Table\,9. Both systems are binaries consisting of a B dwarf and a A giant,  observed at inclination slightly larger than 60\fdg 0.  

The disk contributes 67\% to the flux at the $V$-band in HD\,50526, while it does only 30\% in HD\,170582. This is partly due to bigger secondary star in this system, characterized also by a larger orbital separation and longer photometric periods. Both orbital separation and donor radius are about twice larger in HD\,170582. While the gainer is about 2kK hotter in HD\,170582, the donor is about 2kK coldest.  Interestingly, the full amplitude of variability for the long cycle is 4 times smaller in HD\,170582. In this binary the hot and bright spots are hotter than the surrounding disk compared with those of HD\,50526. However, in terms of absolute temperatures, HD\,50526 have hotter disk spots than HD\,170582, 12785 K (hs) and 11093 K (bs) versus 8980 K and 7899 K, respectively. The disk outer edge temperature is about 9400 K for HD\,50526 and 5400 K for HD\,170582. We notice that the large amplitude long cycle is observed in the system with hotter disk spots. Another peculiar characteristics of HD\,170582 is the presence of a Helium\,5875 absorption feature arising from near the hot spot, not detected in HD\,50526. The presence of Helium absorption suggest high temperature gas around the hot spot, something not observed in the shorter period system. This is in apparent conflict with the temperatures derived from the light curve model, indicating hotter spots in the shorter period system. At present we have not an explanation for this 
discrepancy.

We observe that differences exist in the two systems under comparison, and conjecture that not only orbital and stellar parameters determine the observed behaviour, but probably other variables like the mass transfer rate.

In semidetached Algol, mass from the coldest (donor) star flows through the inner Lagrangian point onto the hotter (gainer) star forming a gas stream. If the gainer is large enough, the stream hits the star forming a hot shock region on its surface. If the gainer is small enough then the stream turns around the star hitting itself and eventually forming an accretion disk when spreading by the effect of the viscosity. This was investigated by 
\citet{1975ApJ...198..383L} who determined this gainer critical radius $r_\mathrm{c}$ as a function of the mass ratio. Actually, due to the finite width of the stream, a range of radii marks the transition boundary between impact systems and disk systems. 
On the other hand the disk, if formed, cannot extend beyond the region when the tidal forces impede its stability, this maximum radius $r_\mathrm{tidal}$ is often also parameterized in terms of the system mass ratio.

\begin{figure}
	\begin{center}
		\includegraphics[trim=0.2cm 0.2cm 0.0cm 0.0cm,clip,width=0.45\textwidth,angle=0]{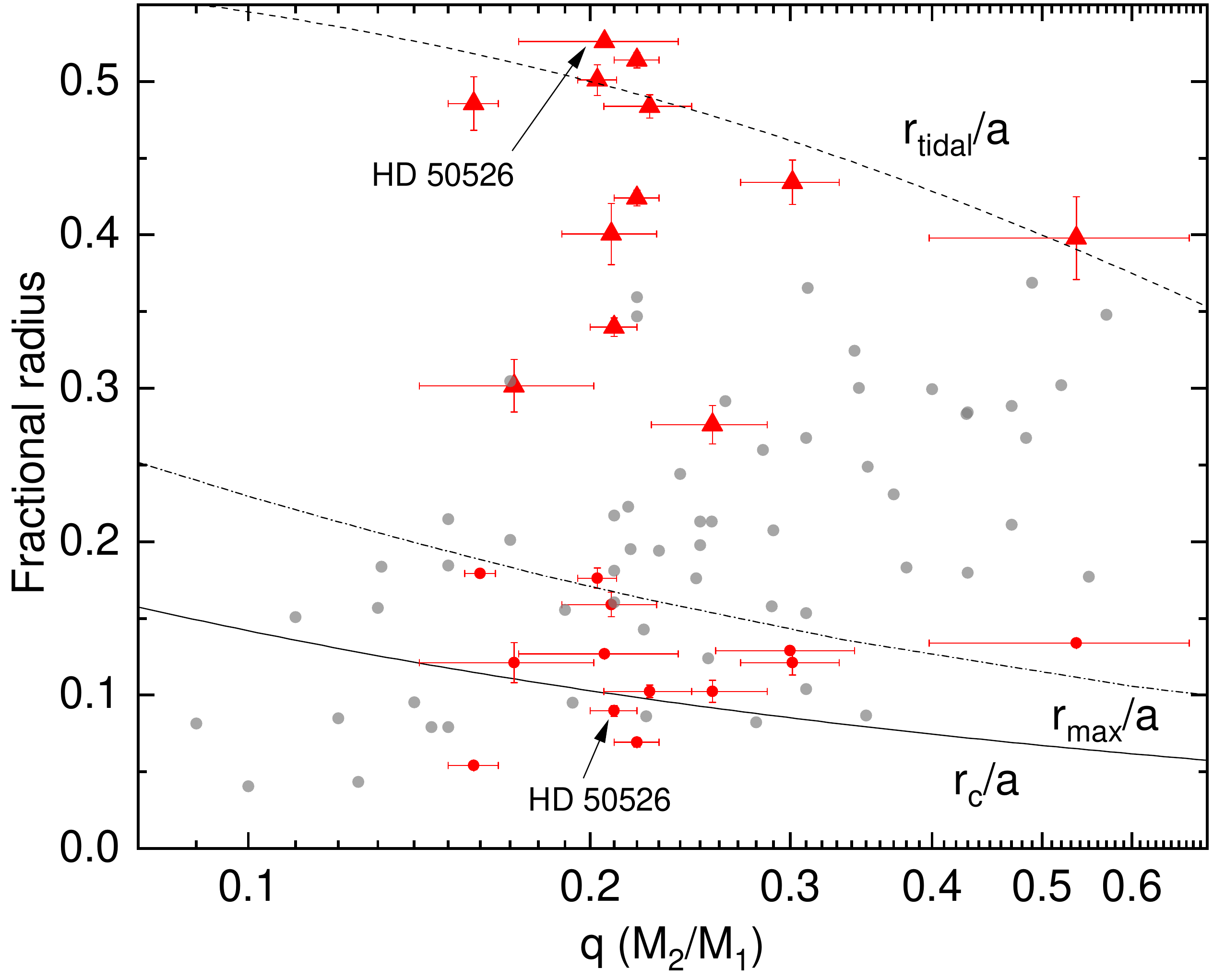}
	\end{center}
	\caption{
		The fractional radius, relative to the orbital separation, for primaries and disks in algols with and without disks. Red circles and triangles show primaries and disks of double periodic variables.
		Semidetached algol primaries from \citet{2010MNRAS.406.1071D} are also shown as black points.
		Below the circularization radius shown by the solid black line, a disk should be formed and below the dashed line, a disk might be formed. The tidal radius indicates the maximum possible disk extension (upper dashed line).  Data are from Table 10.}
	\label{fig:Fig. 14}
\end{figure}

Impact and disk systems can be studied in a diagram of fractional radius (the radius relative to the orbital separation) and mass ratio. We notice that the fractional radius of the hotter star $R_1/a_\mathrm{orb}= 0.127 \pm 0.001$ along with the mass ratio $q= 0.206$ place HD\,50526 into the area of tangential stream impact where most DPVs are found (Figure 14).  This means that if mass transfer occurs due to Roche lobe overflow, the gas stream hits tangentially the more massive star transferring efficiently the angular momentum and eventually spinning-up the gainer star. The fact that most DPVs are found in this area was noticed by \citet{2016MNRAS.455.1728M}. The HD\,50526 disk, on the other hand, has a fractional radius $R_d/a_\mathrm{orb}= 0.526 \pm 0.001$ and is located very near the tidal radius limit, where the influence of tidal forces should disperse the matter particles of the disk limiting its size (Fig.\,\ref{fig:Fig. 14}). This means that the radial extension of the disk is large, since almost fills the Roche lobe of the gainer and material in the outer edge could be subject to tidal forces and flow around the system in non-Keplerian orbits. We notice that not all disks of DPVs reach the tidal limit. Since most radii have been determined for the optically thick disk contributing to the continuum flux and modeling the light curve, still is possible that optically thin parts contributing to line emission extend beyond the measured radius. This could explain the complex morphology of the Doppler maps in HD\,50526.    

We observe a clustering of DPVs around $q$= 0.16-0.3 (Fig.\,14). Only the relatively massive UU\,Cas (total mass $\sim$ 26 M$_{\odot}$) have a larger $q$ $\approx$ 0.54. For this system,  a long cycle of $\sim$ 270 d was detected in the $I_c$-band that needs to be ratified for further photometric campaigns \citep{2020A&A...642A.211M}.

It has been argued that DPV disk luminosities are not accretion-driven but supported by heating from the central star \citep{2016MNRAS.461.1674M}. Can we test if this statement is applicable for HD\,50526? Assuming that the disk is an optically thick, geometrically thin steady-state accretion disk, its temperature is directly related to its mass transfer rate ($\dot{M}$).
We use canonical formulae to determine $\dot{M}= 2.44(2) \times 10^{-6}  ~\mathrm{M_{\odot}\,yr^{-1}}$ \citep[e.g.][Eq. 5.43]{2002apa..book.....F}. If the accretion process is conservative, then an increase in the orbital period of 2.98 s/yr is expected \citep[e.g.][Eq. 4.56]{2001icbs.book.....H}. Considering the error in the orbital period $\epsilon_P = 0.001 ~\mathrm{d}$ and the time baseline of the photometric time series $\Delta T= 2613 ~\mathrm{d}$, we expect at most a drift in the period of $2 \times  \epsilon_P/\Delta T = 24 ~\mathrm{s\,yr^{-1}}$. As we see, more photometric data is needed to constrain further the orbital period and test if the disk has the aforementioned mass transfer rate. Since there are no evidence of significant mass outflows like 
P-Cygni emission profiles, highly shifted lines in radial velocity, strong systemic reddening or a circumstellar nebula, it is possible that the mass transfer in the system is near conservative.  

All these arguments must be evaluated in the context of accretion-decretion disks expected in some DPVs. Since the tangential impact should accelerate the star until critical rotation, we might have a disk with an inner boundary transporting angular momentum and matter outside and the outer boundary transporting mass inside. This kind of relatively massive, self-gravitating accretion-decretion disks are starting to be studied by \citet{2018ApJ...869...19W, 2020CoSka..50..523W} and its importance in the field of algols,
double periodic variables and other close binaries remains to be established by future investigations.

\begin{table}
	\caption{Comparison between the  HD\,50526 data obtained in this paper and those of  HD\,170582  obtained by \citet{2015MNRAS.448.1137M} in the synchronous case. $\Delta V$ is the long-cycle full amplitude and $f_r$ the fractional disk contribution to the $V$ band flux at $\Phi_\mathrm{o}= 0.25$.}
	\label{Tab: Tab. 9}
	\[
	\begin{array}{@{\extracolsep{+0.0mm}}lrr@{}}
	\hline
	\noalign{\smallskip}
	{\rm Quantity} &{\rm HD\,50526} & {\rm HD\,170582}  \\
	\noalign{\smallskip}
	\hline
	\noalign{\smallskip}
	\hline
	{\rm P_o}  [{\rm days}]                 &6.701 &16.871  \\
	{\rm P_l}  [{\rm days}]                 &190 &587 \\
	{\rm P_l/P_o}  &28.4 &34.8 \\ 
	i {\rm [^{\circ}]}                             & 61.3  &67.4 \\
	\Delta V &0.20 &0.05 \\
	a_{\rm orb}  {\rm [R_{\odot}]}    & 28.0&61.2 \\
	{\rm T_d} [{\rm K}]	                   & 9401 & 5410\\
	{\rm T_h} [{\rm K}]			  & 16000&18000 \\
	{\rm T_c} [{\rm K}]			  & 10500&8000 \\
	\cal{R}_{\rm d} {\rm [R_{\odot}]} & 14.7& 20.8 \\
	f_r &0.67 &0.30 \\
	{\rm A_{hs}=T_{hs}/T_d}		 &1.36&1.66 \\
	{\rm \lambda_{hs}}{\rm [^{\circ}]} &343.1&336.6 \\
	{\rm A_{bs}=T_{bs}/T_d}		 &1.18&1.46  \\
	{\rm \lambda_{bs}}{\rm [^{\circ}]} &129.5&134.8\\
	\cal M_{\rm_h} {[\cal M_{\odot}]} & 5.5 & 9.0\\
	\cal M_{\rm_c} {[\cal M_{\odot}]} & 1.1 &1.9 \\
	\cal R_{\rm_h} {\rm [R_{\odot}]} & 3.6   &5.5\\
	\cal R_{\rm_c} {\rm [R_{\odot}]} & 7.1& 15.6 \\
	{\rm log} \ g_{\rm_h}			& 4.1& 3.9 \\
	{\rm log} \ g_{\rm_c}			& 2.8  &2.3 \\
	\hline
	\noalign{\smallskip}
	\end{array}
	\]
\end{table}

\begin{table}
	\caption{Primary star and disk radius relative to the orbital separation and mass ratio for double periodic variables, along with their errors. iDPV stands for OGLE 05155332-6925581 and OGLE157529 for OGLE-BLG-ECL-157529.
		Data are from \citet{2016MNRAS.455.1728M} and references therein and from \citet{2020A&A...642A.211M} and \citet{{2017AstBu..72..321G}} (UU\,Cas), \citet{2020A&A...641A..91M} (OGLE-BLG-ECL-157529, two epochs "a" and "b" are given) and \citet{2018MNRAS.476.3039R} (V495\,Cen). 
	}
	\label{comp}
	\[
	\begin{array}{@{\extracolsep{+0.0mm}}lcccccc@{}}
	\hline
	\noalign{\smallskip}
	{\rm System} &{\rm R_d/a} &{\rm eR_d/a} & {\rm R_1/a} &{\rm eR_1/a} &q &eq \\
	\noalign{\smallskip}
	\hline
	\noalign{\smallskip}
	\hline
	{\rm V393\,Sco }  	        &0.276&	0.012&	0.103&	0.007&	0.256&	0.030\\
	{\rm DQ\,Vel	   }             &0.434&  	0.014&	0.121&	0.008&	0.301&	0.030\\
	{\rm AU\,Mon	      }          &0.302& 	0.017&	0.121&	0.013&	0.171&	0.030\\
	{\rm iDPV		}		&0.401   &0.020&	0.159&	0.008&	0.209&	0.020\\ 
	{\rm \beta\, Lyrae	 }               &0.484& 	0.008&	0.103&	0.004&	0.226&	0.020\\
	{\rm HD\,170582	   }     &0.340&	0.006&	0.090&	0.004&	0.210&	0.010\\ 
	{\rm V495\,Cen	    }             &0.486 &	0.017 &	0.054 &	0.003 &	0.158 &	0.008\\
	{\rm DPV\,065	         }        &0.501 &	0.010 &	0.176 &	0.007 &	0.203 &	0.008\\
	{\rm HD\,50526	      }           &0.526 &	0.001 &	0.127 &	0.001 &	0.206 &	0.033\\ 
	{\rm UU\,Cas	           }      &0.398 &	0.027 &	0.134 &	0.002 &	0.537 &	0.138     \\
	{\rm OGLE\,157529a	    } &0.424&	0.005&	0.069&	0.003&	0.220&	0.010\\
	{\rm OGLE\,157529b	     }&0.514&	0.005&	0.069&	0.003&	0.220&	0.010\\ 
	\noalign{\smallskip}
	\hline
	\end{array}
	\]
\end{table}

\section{Conclusions}
\label{Sec: Sec. 7}
In this work we have studied the double periodic variable HD\,50526 using survey photometry and new spectroscopic data obtained in optical wavelengths with several spectrographs at high and medium resolution. We  summarize our main conclusions as follows :\\

\begin{itemize}
	
	\item  From the photometric study of ASAS data, we determined an orbital period of $P_\mathrm{o} = 6\fd701 \pm 0\fd001 ~\mathrm{d}$ and a long period of $P_\mathrm{l} = 191 \pm 2 ~\mathrm{d} $, both results more reliable than the obtained from spectroscopic analysis due to the larger density of data and time baseline of the photometric time series.
	
	\item We performed a radial velocity study using the \texttt{PIKAIA} code resulting in the best orbital parameters given in Table\,\ref{Tab: Tab. 5}. In particular, we obtain a half-amplitude of radial velocity for the cold star $K_{2} = 153 ~\mathrm{km\,s^{-1}}$ and a systemic velocity $\gamma = -5.17 ~\mathrm{km\,s^{-1}}$.

	\item Sharp lines of the cold stellar component dominate the spectrum. A comparison with a grid of synthetic stellar spectrum indicates a temperature $T_\mathrm{2}= 10500 \pm 125 ~\mathrm{K}$, surface gravity $\log{g}=3.0 \pm 0.25 ~\mathrm{dex}$, 
	and $v\sin{i}= 70 \pm 20 ~\mathrm{km\,s^{-1}}$.
	
	\item Double emission flanks surrounding the H$\alpha$ cold star absorption lines are detected, suggesting the existence of an accretion disk in the system. 
	
	\item We find, through light curve modeling, a complete solution for the system including stellar and disk parameters that are listed in Table\,\ref{Tab: Tab. 8}. We find a binary consisting of a pair of  B5\,V + A0\,III stars. The cold star fills its Roche lobe and transfer mass onto the hot star, forming and accretion disk of radius $14.7 ~\mathrm{R_\odot}$ and outer edge temperature 9400\,K. 
	
	\item According to the model, the accretion disk has two bright and hot active regions, 1.36 and 1.18 times hotter than the surrounding disk, and located 343 and 47 degrees apart from a line connecting the stellar centers as measured in the orbital motion direction.

	\item We construct Doppler maps for the H$\alpha$ and H$\beta$ lines, finding additional and complementary evidence of circumstellar emitting material located around the gainer star, consistent with the existence of the accretion disk. 
	
	\item The long-cycle variability, although unknown in its nature, could be related to the existence of mass flows in the systems as shown in the Doppler maps and general spectroscopic variability.

\end{itemize}


\section*{Acknowledgements}

Part of the observations presented in this paper were conducted by the late Zbigniew Ko\l{}aczkowski during his postdoctoral stay in the University of Concepci\'on, Chile. We thanks Sebasti\'an Otero for useful insights on photometric data surveys. 
This publication makes use of VOSA, developed under the Spanish Virtual Observatory project supported from the Spanish MICINN through grant AyA2011-24052. J.R. and R.E.M. gratefully acknowledges support from the Chilean BASAL Centro de Excelencia en Astrofísica y Tecnologías Afines (CATA) grant PFB-06/2007, AFB-170002 and support by VRID-Enlace 218.016.004-1.0. This research was funded in part by a scholarship from Faculty of Physical Sciences and Mathematics of the Universidad de Concepción (UdeC). R.E.M. and D.R.G.S. thanks for funding through FONDECYT regular, project codes 1201280 and 1190621, and through the \textquotedblleft Concurso Proyectos Internacionales de Investigacion, Convocatoria 2015\textquotedblright, project code PII20150171. G.\,D. gratefully acknowledge the financial support of the Ministry of Education, Science and Technological Development of the Republic of Serbia through contract No. 451-03-68/2020-14/200002. S. Z. acknowledges PAPIIT-DGAPA-UNAM grant IN102120. I.A. and M.C. acknowledges support from FONDECYT project Nº 1190485. I. A. is also grateful for the support from FONDECYT project Nº 11190147. Also this work has made use of data from the European Space Agency (ESA) mission {\it Gaia} (\url{https://www.cosmos.esa.int/gaia}), processed by the {\it Gaia} Data Processing and Analysis Consortium (DPAC, \url{https://www.cosmos.esa.int/web/gaia/dpac/consortium}). Funding for the DPAC has been provided by national institutions, in particular the institutions participating in the {\it Gaia} Multilateral Agreement. 


\end{document}